\documentclass[aps,pra,twocolumn,groupedaddress,nofootinbib]{revtex4-2}

\usepackage{graphicx,amsmath,hyperref}

\begin{document}

\title{An atomic Faraday beam splitter for light generated from pump degenerate four-wave mixing in a hollow-core photonic crystal fiber}

\author{Ioannis Caltzidis, Harald K{\"u}bler, Tilman Pfau, Robert L{\"o}w and Mark A. Zentile}
\altaffiliation[Present address: ]{OnePlanet Research Center, Bronland 10, 6708 WH Wageningen, The Netherlands}
\affiliation{5. Physikalisches Institut and Center for Integrated Quantum Science and Technology, University of Stuttgart, Pfaffenwaldring 57, D-70569 Stuttgart, Germany}

\date{\today}

\begin{abstract}
We demonstrate an atomic Faraday dichroic beam splitter suitable to spatially separate signal and idler fields from pump degenerate four-wave mixing in an atomic source. By rotating the plane of polarization of one mode $90^{\circ}$ with respect to the other, a subsequent polarizing beam splitter separates the two frequencies, which differ by only 13.6 GHz, and achieves a suppression of $(-26.3\pm0.1)$ and $(-21.2\pm0.1)$\,dB in the two outputs, with a corresponding transmission of 97 and 99\,\%. This technique avoids the need to use spatial separation of four-wave mixing modes and thus opens the door for the process efficiency to be enhanced in waveguide experiments. As a proof-of-principle we generate light via four-wave mixing in $^{87}$Rb loaded into a hollow-core photonic crystal fiber and interface it with the atomic Faraday dichroic beam splitter.
\end{abstract}

\maketitle

\section{Introduction}\label{sec:intro}

Non-linear optical effects are well known to be greatly enhanced when guided through optical fibers, due to both high-intensities and large interaction lengths which far exceed the Rayleigh length~\cite{Benabid2005}. The use of hollow-core photonic crystal fibers (HC-PCFs)~\cite{Fevrier2011} conveniently allows the light mode to interface with a gas \cite{Debord2019} or liquid medium~\cite{Horan2012} of choice. When loaded with a highly resonant medium, such as an atomic gas, the HC-PCF system shows pronounced non-linear effects at low powers. Electromagnetically induced transparency \cite{Ghosh2006a}, four-wave mixing \cite{Londero2009}, cross-phase modulation~\cite{Perrella2013a}, photon memories \cite{Sprague2014} and singe-photon sources~\cite{Cordier2020} have all been shown to be enhanced with an atomic gas loaded HC-PCF system.

In recent years four-wave mixing (FWM) in atomic vapours has received considerable attention because it allows the generation of narrowband non-classical light~\cite{McCormick2006,MacRae2012,Shu2016,Whiting2017,Ripka2018} which is naturally compatible with other atom-based applications such as single-photon storage~\cite{Julsgaard2004,Hosseini2011,Michelberger2015a}. Pump degenerate FWM~\cite{McCormick2006,MacRae2012,Podhora2017}, where the two pump photons have the same frequency,  allows for a simplified experimental setup because only one laser is needed to pump the medium. Therefore, pump-degenerate FWM should be more applicable as a quantum technology. However, to achieve a high spectral brightness in a free-beam configuration, high-power solid state lasers are used~\cite{MacRae2012} which are cumbersome and expensive. A boost in efficiency will be required to allow simple low-power diode lasers to take over the role as the pump and push the system into the realm of practical quantum technology. For this HC-PCFs could be used, since they allow the FWM effect to occur over a much increased path length, and the efficiency scales with this length squared~\cite{Bratfalean1996}.

The hollow-core fiber system, however, presents a challenge for filtering the different frequencies since the signal, idler and pump fields are all spatially overlapping. The pump field is cross-polarized with respect to the signal and idler fields, however a subsequent polarizer will not sufficiently remove the pump field, additional methods are required~\cite{MacRae2012,Palittapongarnpim2012}. In this paper we show that by using isotopically pure $^{87}$Rb as the FWM medium, the pump field can be removed from the output by using a subsequent isotopically pure $^{85}$Rb cell which is resonant with the pump. The $^{85}$Rb cell is thus used as an ultra-narrow notch filter which allows the signal and idler fields to pass. This leaves the challenge of separating the signal and idler fields from each other. Since these fields have the same polarization, their frequency difference (just twice the ground state hyperfine splitting) is the only degree of freedom with which to separate them. We show that the Faraday effect in another atomic vapour can be used as the beam splitter. We achieve the separation of signal and idler modes by rotating the plane of polarization of one mode by 90$^\circ$ with respect to the other.

The Faraday effect in atomic vapours can be strong. High Verdet constants have being demonstrated~\cite{Weller2012d} with minimal absorption. Given that this effect is highly frequency dependent and strongest when close to an atomic transition, it is commonly used to create ultra-narrow bandpass filters (see ref.~\cite{Gerhardt2018} and references therein). However, only a few examples have been shown where the Faraday effect was used as a beam splitter \cite{Abel2009,Portalupi2016}. In this study we show that computational optimisation can be used to design the Faraday dichroic beam splitter (FDBS), and we experimentally show that it achieves unprecedented transmission and purity in both output ports. 

Furthermore, as a proof-of-principle, we describe a pump-probe experiment where light generated from $^{87}$Rb atoms loaded into a HC-PCF was interfaced with the FDBS. We show that we can use the FDBS to estimate the ratio of FWM to Raman processes responsible for the gain seen in the pump-probe experiment. 

\section{Theory and background}

\begin{figure}
	\includegraphics[width=\linewidth]{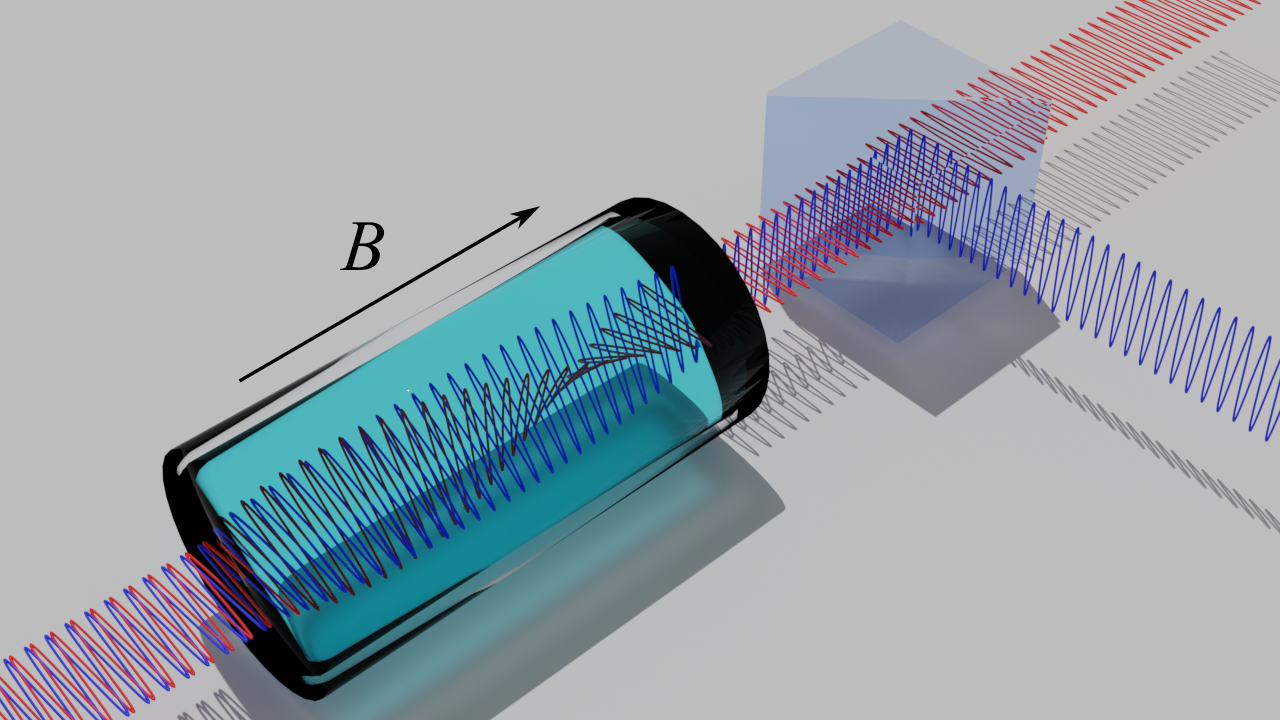}
	\caption{Illustration of an atomic Faraday dichroic beam splitter. Two light waves of with slightly different frequencies are initially polarized in the same direction before being incident on an atomic vapour cell. An axial magnetic field ($B$) in the atomic vapour induces a Faraday rotation of the plane of polarization, which differs in strength for the two frequencies of light. A $90^\circ$ difference in the Faraday rotation for the two frequencies allows them to be perfectly separated at a polarizing beam splitter cube.}
	\label{fig:setup0}
\end{figure}

The concept of an FDBS is illustrated in Fig.~\ref{fig:setup0}. Two different frequencies of light, with the same polarization, are incident on an atomic vapour cell. An axial magnetic field ($B$) causes circular birefringence in the atomic medium, which in turn causes a rotation of the plane of polarization (the Faraday effect \cite{Faraday1846}). Due to the strong frequency dependence of the circular birefringence, the two frequencies of light will have their plane of polarization rotated by different amounts. If the difference in polarization after the vapour cell is $90^\circ$ then the two frequencies can be perfectly separated at a polarizing beam splitter. It should be noted that the magnetic field will also induce differential absorption between the circular components of linearly polarized light (circular dichroism), which means the light emerging from the cell cannot remain entirely linearly polarized, but rather elliptically polarized. Therefore perfect separation is impossible with linear polarizing beam splitters, however, since absorption decreases with detuning from resonance faster than dispersion \cite{Siddons2009}, the effect of circular dichroism can be minimised while keeping a substantial circular birefringence. It is for this reason that it is possible for the light to remain almost linearly polarized while exhibiting significant Faraday rotation.

The challenge is to engineer the properties of the atomic vapour such that the difference in Faraday rotation is as close to $90^\circ$ as possible while maximising transmission at the two frequencies.
In a thermal atomic medium, the transmission and dispersion spectra can be modelled very accurately in the weak-probe regime \cite{Weller2012}. Simple analytical expressions for the Faraday rotation signals can be used when far off resonance \cite{Wu1986,Kemp2011}, since Doppler broadening and hyperfine structure can be ignored \cite{Siddons2009}. However, when on or near resonance, simple analytical expressions cannot be used due to the large number of partially overlapping transitions of varying strengths, which create a complicated spectrum. We use a well established model, that has proven to be highly accurate for predicting absorption and dispersion spectra for alkali-metal vapours~\cite{Siddons2008,Weller2011,Weller2012}. To compute the model have chosen to use the open source \mbox{\emph{ElecSus}} program~\cite{Zentile2014a,Keaveney2018}. A description of the underlying model is given in ref.~\cite{Zentile2014a}.

Various experimental parameters such as the strength of the $B$-field, it's orientation with respect to the light propagation axis, cell temperature ($T$), cell length, and angle of the polarizers will all affect the output of the dichroic beam splitter device. In principle all of these parameters can be optimised by computational means \cite{Kiefer2014,Zentile2015,Keaveney2018d}. To decrease computation time we choose to only keep $B$ and $T$ as parameters and fix all the other degrees of freedom. Furthermore, we consider only an axial magnetic field, and also only the case where the final polarizer is aligned to the input light polarization for maximum transmission (defined here as the $x$-direction). The length of the vapour is fixed to 2\,mm because this length is short enough to allow high-strength, uniform magnetic fields to be achieved conveniently with permanent magnets (up to $\sim$\,10\,kG~\cite{Reed2018}).

\section{Design by computational optimisation}

Effective optimisation is achieved by careful design of a cost function which captures the deviation from optimal working conditions. This cost function is then minimised while changing the magnetic field and temperature to find these optimal working parameters. High transmission and high purity in each output is desired, and as such the following cost function is used:
\begin{equation}
\begin{aligned}
C=\left\{\begin{array}{ll}C_1 =&\left(I_{y}\left(\nu_{1}\right)+I_{x}\left(\nu_{2}\right)\right)^2\\
\hspace{5mm}&-\left(I_{x}\left(\nu_{1}\right)+I_{y}\left(\nu_{2}\right)\right)^2, C_1 < C_2 \\
C_2 =&\left(I_{x}\left(\nu_{1}\right)+I_{y}\left(\nu_{2}\right)\right)^2\\
\hspace{5mm}&-\left(I_{y}\left(\nu_{1}\right)+I_{x}\left(\nu_{2}\right)\right)^2, C_2 < C_1 \end{array} \right.
\label{eq:cost_1}
\end{aligned}
\end{equation}
where $I_x$ and $I_y$ are the components of the light intensity exiting the vapour cell of orthogonal linear polarizations, and $\nu_1$ and $\nu_2$ are the two frequencies of interest. Note that if there is either a complete loss of light or no frequency discrimination (i.e $I_{x,y}(\nu_{1,2})=I_{y,x}(\nu_{1,2})$) the cost function reaches a maximum value of zero. The minimum value of -4 is only achieved when there is both full transmission and complete separation of the two frequencies. Also, since it is unimportant from which output the two frequency components emerge, the cost function is evaluated for both cases ($C_1$ and $C_2$) with the smaller taken.
\begin{figure}
	\includegraphics[width=\linewidth]{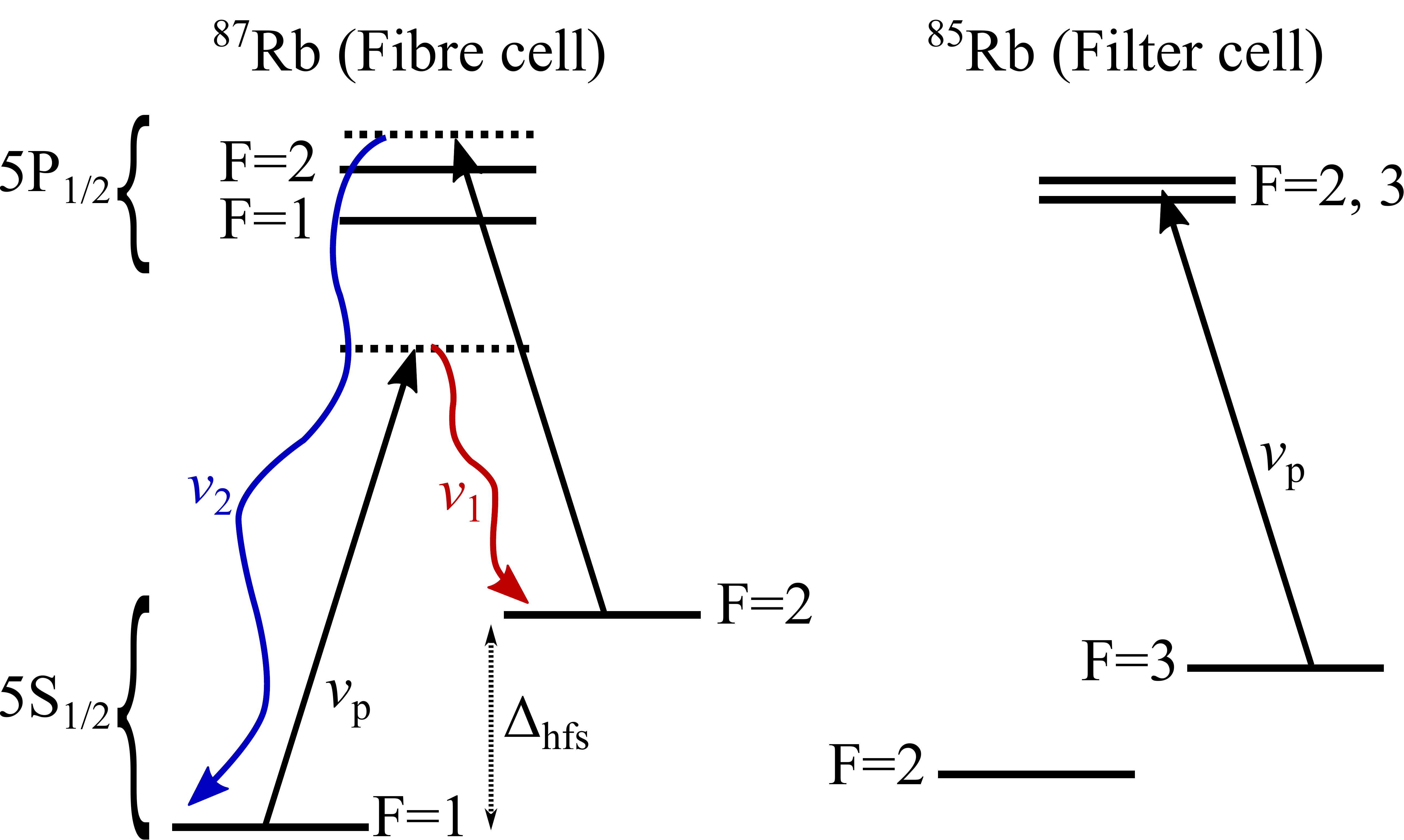}
	\caption{(Left) Double-lambda energy level scheme to produce photon pairs via four-wave mixing in $^{87}$Rb. Two pump photons at frequency $\nu_p$, are converted into a pair photons of frequency $\nu_1 = \nu_p - \Delta_{\mathrm{hfs}}$ and $\nu_2 = \nu_p + \Delta_{\mathrm{hfs}}$. (Right) The pump frequency is be tuned to be on resonance with the Doppler broadened $F=3 \rightarrow F'=2,3$ transition in $^{85}$Rb, allowing the pump to be filtered }
	\label{fig:leveldiagram}
\end{figure}

The frequencies $\nu_1$ and $\nu_2$ were chosen for this study by considering the case of light generated in $^{87}$Rb gas from a double-lambda scheme, as shown in figure~\ref{fig:leveldiagram}. A pump laser is blue detuned approximately 0.8 GHz detuned from the $F=2 \rightarrow F'=2$ transition on the D$_1$ line. This scheme has been chosen because it is close enough to resonance to show significant emission at the frequencies $\nu_1$ and $\nu_2$, which complete the double lambda scheme, while allowing the pump light to be eliminated by subsequent cell containing isotopically pure $^{85}$Rb. This results in the frequencies to separate being at detuning values of -8.23 and 5.43 GHz from the Rb D$_1$ global linecentre of 377.107407 THz \cite{Zentile2014a}.
\begin{figure}
	\includegraphics[width=\columnwidth]{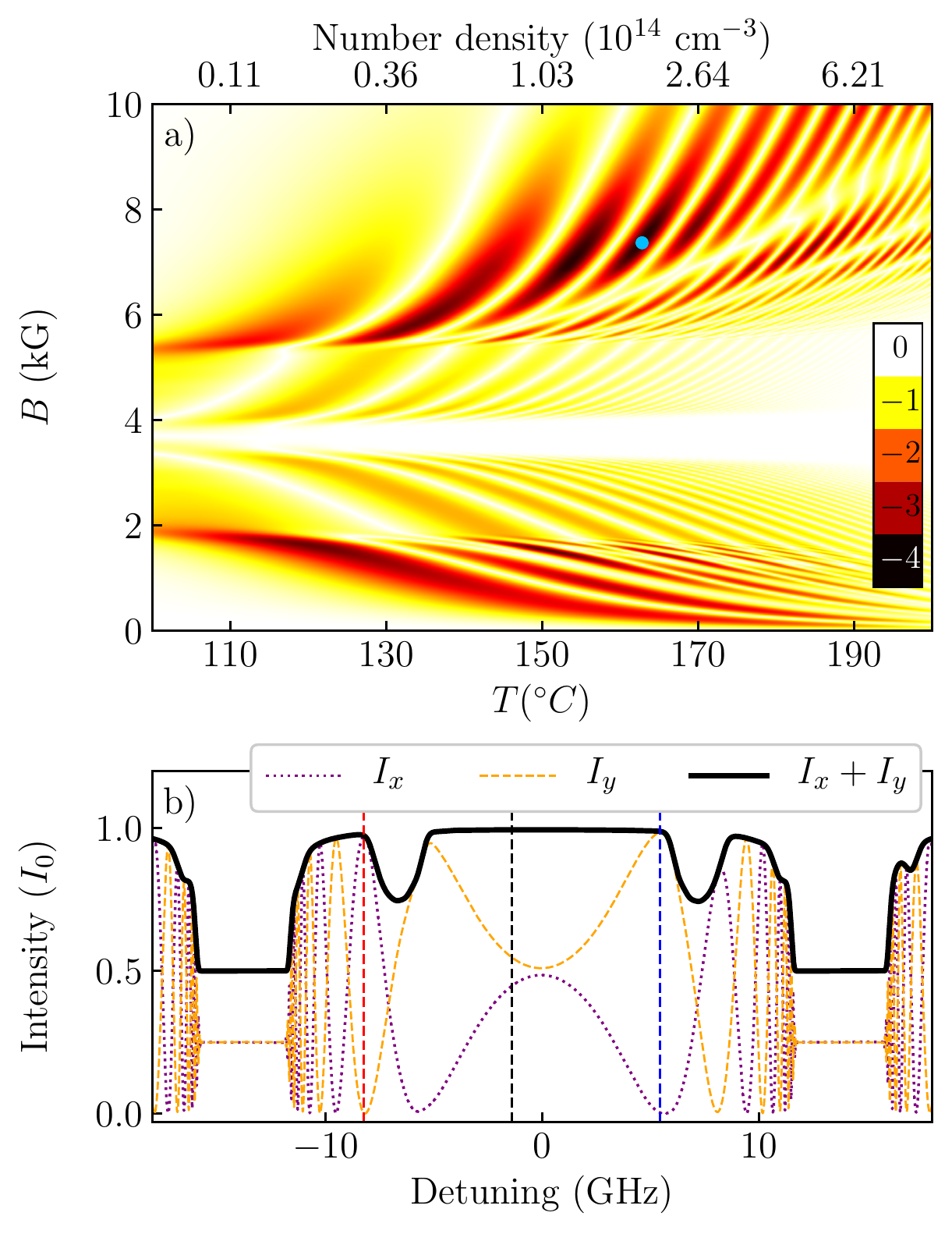}
	\caption{a) Cost function (eq.~\ref{eq:cost_1}) for a 2 mm long Rb vapour cell (isotopic ratio: 99.84~\% $^{85}$Rb, 0.16~\% $^{87}$Rb) mapped over magnetic field strength ($B$) and temperature ($T$) in steps of 10\,G and 0.1\,$^\circ$C respectively. The solid green circle shows the location of the global optimal solution of -3.84 at 7.366 kG and 162.8~$^\circ$C. b) Intensity spectra, corresponding to the global optimal solution, as a function of global detuning in units of the initial intensity ($I_0$). The black solid curve shows transmission while the purple dotted and orange dashed curves show its components in the $x$- and and $y$-direction respectively. The dashed vertical lines (from left to right) show the frequencies $\nu_1$, $\nu_p$ and $\nu_2$ according to figure~\ref{fig:leveldiagram}.}
	\label{fig:compOpt}
\end{figure}

The cost function is efficiently minimised by employing a carefully constructed global fitting routine \cite{Zentile2015}. However, to illustrate the structure of the cost function we have chosen to map it in detail over the a large parameter range, from 100 to 200~$^\circ$C and from 0 to 10~kG. The result for an isotopically pure $^{85}$Rb cell is shown in panel a) of figure~\ref{fig:compOpt}. There are many minima seen emphasising the need for a global optimisation routine. A key feature of the map is that the minima are somewhat periodic in number density which is simply explained by solutions occurring for Faraday rotation angles ($\phi$) that occur at approximately $n\frac{\pi}{2}$ for one frequency and $(n+1)\frac{\pi}{2}$ for the other. Another feature is that minima are seen for two magnetic field domains, $<2\,$kG and $>5\,$kG. In the lower field domain the magnetic shift of the atomic lines is small and as such the atomic transitions occur between $\nu_1$ and $\nu_2$. In the higher field regime the magnetic shift is now such that the strong atomic transitions lie outside of the region between $\nu_1$ and $\nu_2$. In both these regimes it is possible to achieve a near 90$^\circ$ differential Faraday rotation with little absorption. However, in between these magnetic field regimes the atomic transitions coincide with one or more of the frequencies of interest, causing absorption and creating ellipticity in the output light, and therefore increasing the cost function.

The global optimal solution ($C = -3.84$) is found at a cell temperature of 162.8$\,^\circ$C and magnetic field of $7.366\,$kG. At this high magnetic field the atoms are in the Hyperfine Paschen-Back regime \cite{Sargsyan2012a,Weller2012d,Weller2012c,Zentile2014,Sargsyan2015a}. The transmission spectrum of the vapour and its polarization components in the $x$- and $y$-direction are shown in panel b) of figure~\ref{fig:compOpt}. The spectra are almost symmetrical around the weighted linecentre, a common feature of the hyperfine-Paschen Back regime. Also, at this high temperature the vapour is highly circularly dichroic at detunings of approximately $\,15$~GHz, being optically thick for one circular polarization while being transparent for the other. This results in transmission values of approximately 0.5 in these regions with the resulting $I_x$ and $I_y$ intensities being almost 0.25. Panel b) of figure~\ref{fig:compOpt} also clearly shows high transmission into the $I_x$ channel for $\nu_1$ $(97~\%)$ and high transmission into the $I_y$ channel for $\nu_2$ $(99~\%)$.

\section{Experimental results}

To experimentally test the solution found by computational optimisation, the FDBS was constructed and tested with weak probe laser spectroscopy. A 2\,mm long vapour cell was surrounded with two permanent ring magnets to apply large magnetic field. The magnets were custom designed to give a near uniform magnetic field which matched the optimal value found. Due to manufacturing tolerance, subsequent small adjustments of the magnetic field is needed to match the magnetic field. This was achieved by both adjusting the magnets' separation and their temperatures. A weak probe beam from an external cavity diode laser was then sent through a polarizing beam splitter cube (PBS) before passing through the vapour cell and then being incident on a Brewster-cut Glan-Taylor polarizing beam splitter (BC-GT). The intensity of the transmitted ($I_x$) and reflected ($I_y$) beams from the BC-GT were then measured with amplified photodetectors. Before heating the vapour cell, the PBS and BC-GT were aligned such that the laser power was maximised through the transmission port of the BC-GT and minimised at its reflection port. The use of the BC-GT as an analyzer improves the polarization purity in the reflected port by reducing reflection of p-polarized light. When measured with an optically thin, room-temperature vapour cell, we found the extinction ratios for the whole setup between the two polarizers to be $(-47.10\pm0.06)\,$dB and $(-38.041\pm0.006)\,$dB for the $I_x$ and $I_y$ outputs respectively.

\begin{figure}
	\includegraphics[width=\columnwidth]{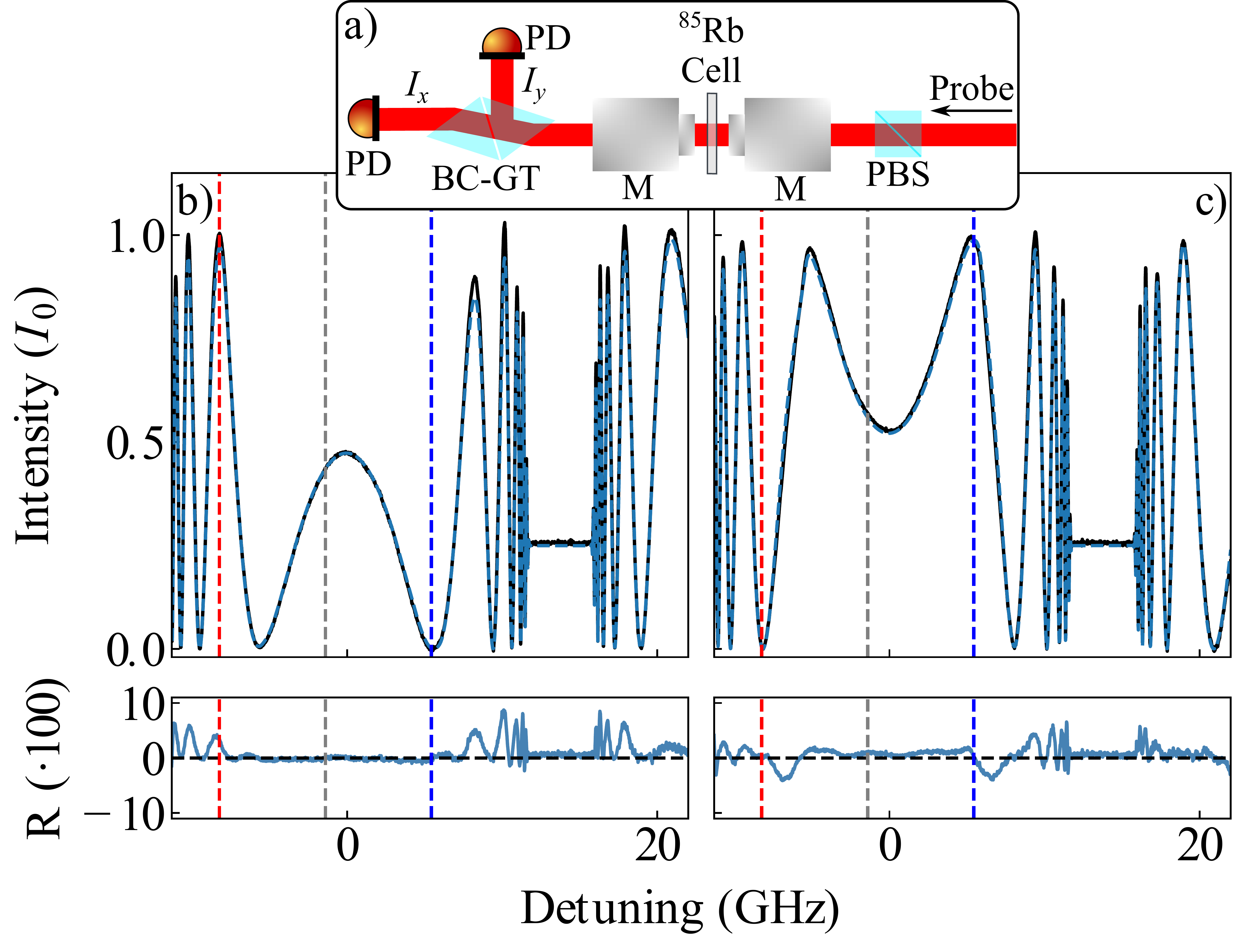}
	\caption{a) Schematic of the apparatus for weak probe laser spectroscopy of the Faraday dichroic beam splitter. A pair of permanent ring magnets (M) surround a $^{85}$Rb vapour cell giving a strong axial magnetic field. The probe laser passes through a polarizing beam splitter (PBS) and the vapour cell before being incident on a brewster-cut Glan-Taylor polarizing beam splitter (BC-GT) which separates the light into orthonal polarization components, $I_x$ and $I_y$, before being detected by photodetectors (PDs). Example intensity spectra in units of the input intensity ($I_0$) are shown for $I_x$ and $I_y$ in panels b) and c) respectively. The solid black lines show the experimental data while the dashed blue lines show theory fits with the magnetic field and cell temperature as parameters. Below each plot the residuals (R) between experiment and theory are plotted. The dashed vertical lines (from left to right) show the frequencies $\nu_1$, $\nu_p$ and $\nu_2$ according to figure~\ref{fig:leveldiagram}.}
		\label{fig:FaradaySpectroscopy}
\end{figure}

The cell was then heated. After the setup reached thermal equilibrium the laser was scanned and five spectra were recorded in quick succession. An example spectrum is visible in figure~\ref{fig:FaradaySpectroscopy}, along with a fit using ElecSus~\cite{Zentile2014a,Keaveney2018}. We see good agreement between experiment and theory with an RMS deviation of 2\,\%. From fitting the five spectra we find a magnetic field of $\left(7.362\pm0.001\right)$\,kG and a temperature of $\left(162.86\pm0.01\right)\,^\circ$C, where the uncertainties represent the standard deviation in the fit parameters over the five spectra.
The difference between the optimal parameters and these measured ones are very small and should not cause any significant degradation in performance.

\begin{figure}
	\includegraphics[width=\columnwidth]{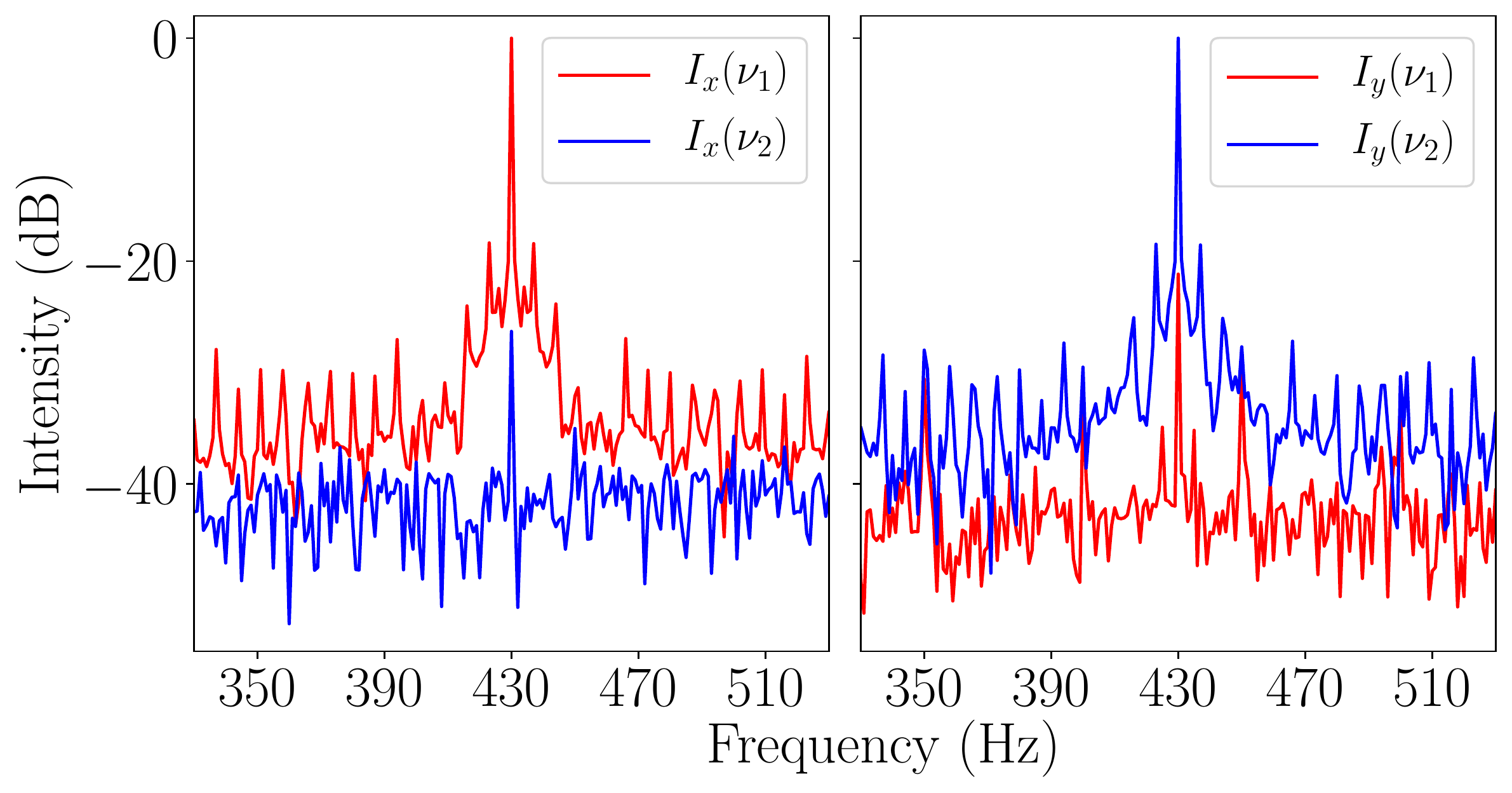}
	\caption{Normalised intensity as a function of frequency while modulating the probe beam at 430 Hz. The left(right) panel shows the intensity measured at the $I_x$($I_y$) port of the beam splitter, while the red(blue) curves correspond to the case where the laser frequency was set to $\nu_1$($\nu_2$).}
	\label{fig:IntDb_vs_freq}
\end{figure}

These results experimentally confirm that we achieve high transmission into $I_x$ at $\nu_1$, and high transmission into $I_y$ at $\nu_2$ as predicted. Using the fitted theoretical spectra we can also predict the spectral purity in each channel by calculating the extinction ratios,
\begin{equation}
\begin{aligned}
r_x &= I_x\left(\nu_2\right)/I_x\left(\nu_1\right) \\
r_y &= I_y\left(\nu_1\right)/I_x\left(\nu_2\right),
\label{eq:ExtinctionRatios}
\end{aligned}
\end{equation}
which were found to be $-27.9$ and $-24.2\,$dB for $r_x$ and $r_y$ respectively. Note that these extinction ratios are calculated assuming perfect polarizers and no birefringence from the cell windows. For the experimental measurements however, our laser spectroscopy method is not precise enough to give a useful comparison with theory. For this measurement, we instead use a technique similar to lock-in amplification, where the laser frequency is fixed to one of the frequencies of interest ($\nu_1$) while modulating the laser with an optical chopper. Using a Fourier transform we take the signal power at the chopper frequency only and thereby remove the effect of noise at all other frequencies. In this way we get a more precise measure of $I_x\left(\nu_1\right)$ and $I_y\left(\nu_1\right)$. Repeating the process at the other frequency of interest gives a measurement of $I_x\left(\nu_2\right)$ and $I_y\left(\nu_2\right)$. Figure~\ref{fig:IntDb_vs_freq} shows the result of this measurement. From this data we calculate $r_x = (-26.3\pm0.1)\,$dB, $r_y = (-21.2\pm0.1)\,$dB, where the uncertainties are derived from the uncertainty in the change in laser power after changing its frequency. These extinction ratios are smaller in magnitude than the theoretical best values which could be due to a small drift in the cell temperature during the time taken to change the laser frequency. Nevertheless, for the purpose of heralded single photon sources, the residual impurity would only negligibly decrease the heralding efficiency.

\section{Separation of light from four-wave mixing in a HC-PCF}

In this section a proof-of-principle experiment is described where the Faraday beam splitter is used to separate light generated by seeded four-wave mixing from thermal $^{87}$Rb atoms in a HC-PCF. We use a kagome structured hollow-core photonic crystal fiber with a $60\,\mu$m core diameter and a length of 10\,cm. More details about the fibre used are given in ref.~\cite{Epple2014}. The fiber is fully encapsulated within a borosilicate glass cell containing isotopically purified Rb vapour (isotopic ratio: 98.8~\% $^{87}$Rb, 1.2~\% $^{85}$Rb). Fiber-cells of this type show fast fiber filling times at tens of minutes~\cite{Gutekunst2016,Gutekunst2017} compared to fibers mounted in steel based vacuum chambers which can take months to fill~\cite{Epple2014}. The fiber-cell system also benefits from having a persistent high optical depths without light induced atomic desorption, which has been exploited for steel-based vacuum chamber systems~\cite{Kaczmarek2015a,Donvalkar2015a}.

 \begin{figure}
	\includegraphics[width=\columnwidth]{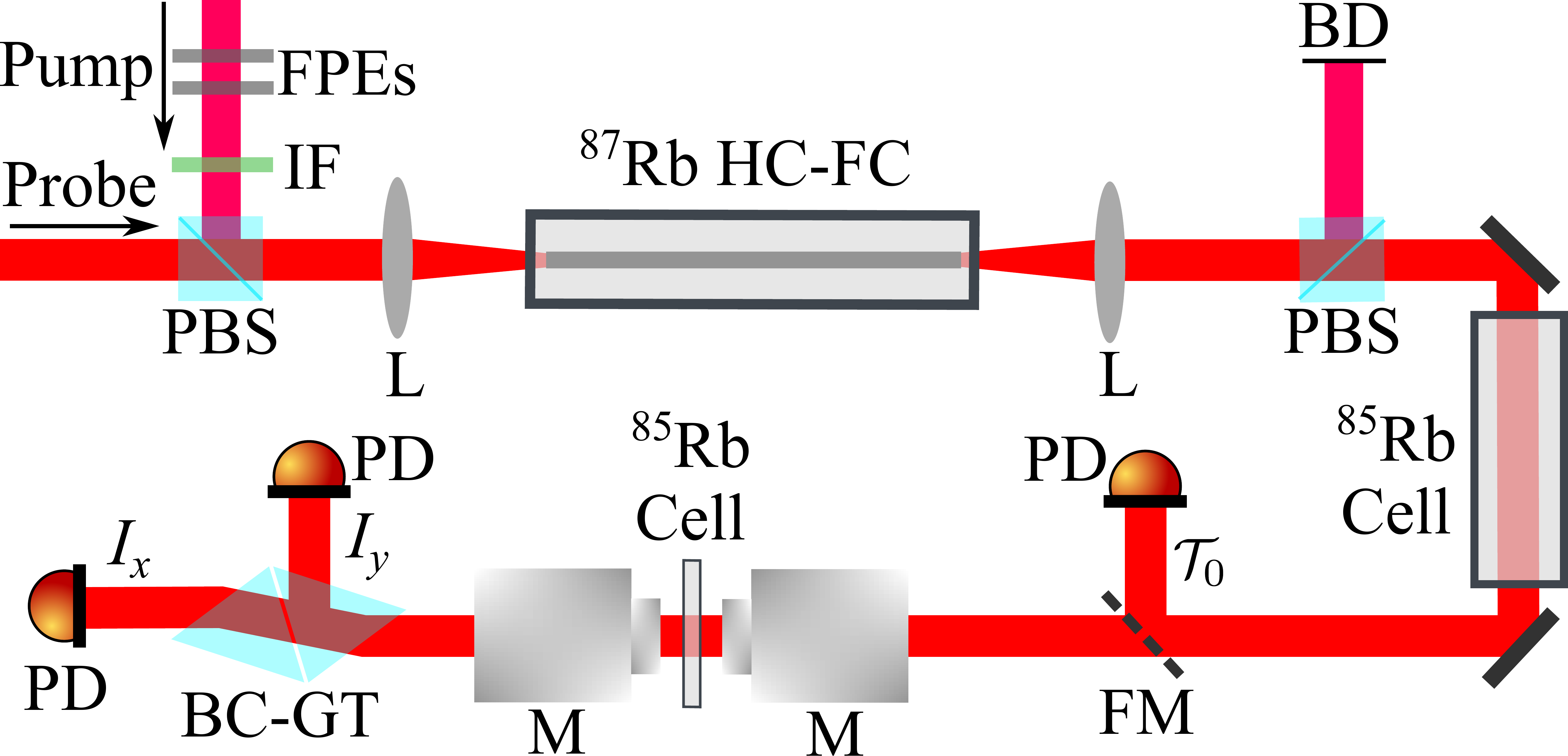}
	\caption{Experimental apparatus for separation of signal and idler modes produced via four-wave mixing in a Rb loaded HC-PCF. The pump beam is spectrally purified by passing though the two solid Fabry-P\'erot etalons (FPEs) and an interference filter before being focussed with a lens (L) into the hollow-core fiber mounted within an isotopically pure $^{87}$Rb vapour cell. The orthogonally polarized probe laser is overlapped with the pump in the fiber. The hollow-core fiber cell (HC-FC) is heated to approximately $70^\circ$C. A polarizing beam splitter (PBS) cube after the HC-FC removes most of the pump light to a beam dump (BD) while transmitting the probe. A subsequent 75 mm long isotopically pure $^{85}$Rb cell, heated to approximately $75^\circ$C, then scatters any remaining pump light. A flipper mirror (FM) can be used to direct the beam onto a photodetector (PD) to measure the probe transmission at this point ($\mathcal{T}_0$) or direct the probe to the Faraday dichroic beam splitter.}
	\label{fig:setup1}
\end{figure}
A diagram of the experimental arrangement is shown in figure~\ref{fig:setup1}. An external cavity diode laser (ECDL) is used for the pump beam and is fixed to be resonant with the Doppler broadened $F=3\rightarrow F^\prime=2,3$ transition in $^{85}$Rb. The pump passes through two solid etalons and an interference filter which serve to greatly suppress broadband emission from the laser diode which is outside of the lasing wavelength. The probe beam comes from another ECDL and is scanned approximately $25\,$GHz around the D$_1$ weighted linecentre. The pump beam is overlapped with the probe beam inside the hollow-core fiber. The pump and probe have a power of 1.5\,mW and $0.1\,\mu$W inside the fiber. The two beams are orthogonally polarized and a polarizing beam splitter (PBS) cube after the fiber cell transmits the probe while removing approximately 99\,\% of the pump light. The pump at this point is still over two orders of magnitude stronger at than the signal and idler fields, and so a heated 75\,mm long $^{85}$Rb cell is used to eliminate the pump. We detected no remaining pump light even when using single photon counting modules, and from these measurements we estimate the pump light to have a power of less than $1\,$fW. Also we estimate the loss on the signal and idler fields, due to imperfect AR-coatings and scattering in the atomic vapour, to only be less than $1\,\%$. After passing the 75\,mm long $^{85}$Rb cell, a flipper mirror can then be used to measure the probe at this point or direct the beam to the FDBS setup.

\begin{figure}
	\includegraphics[width=\columnwidth]{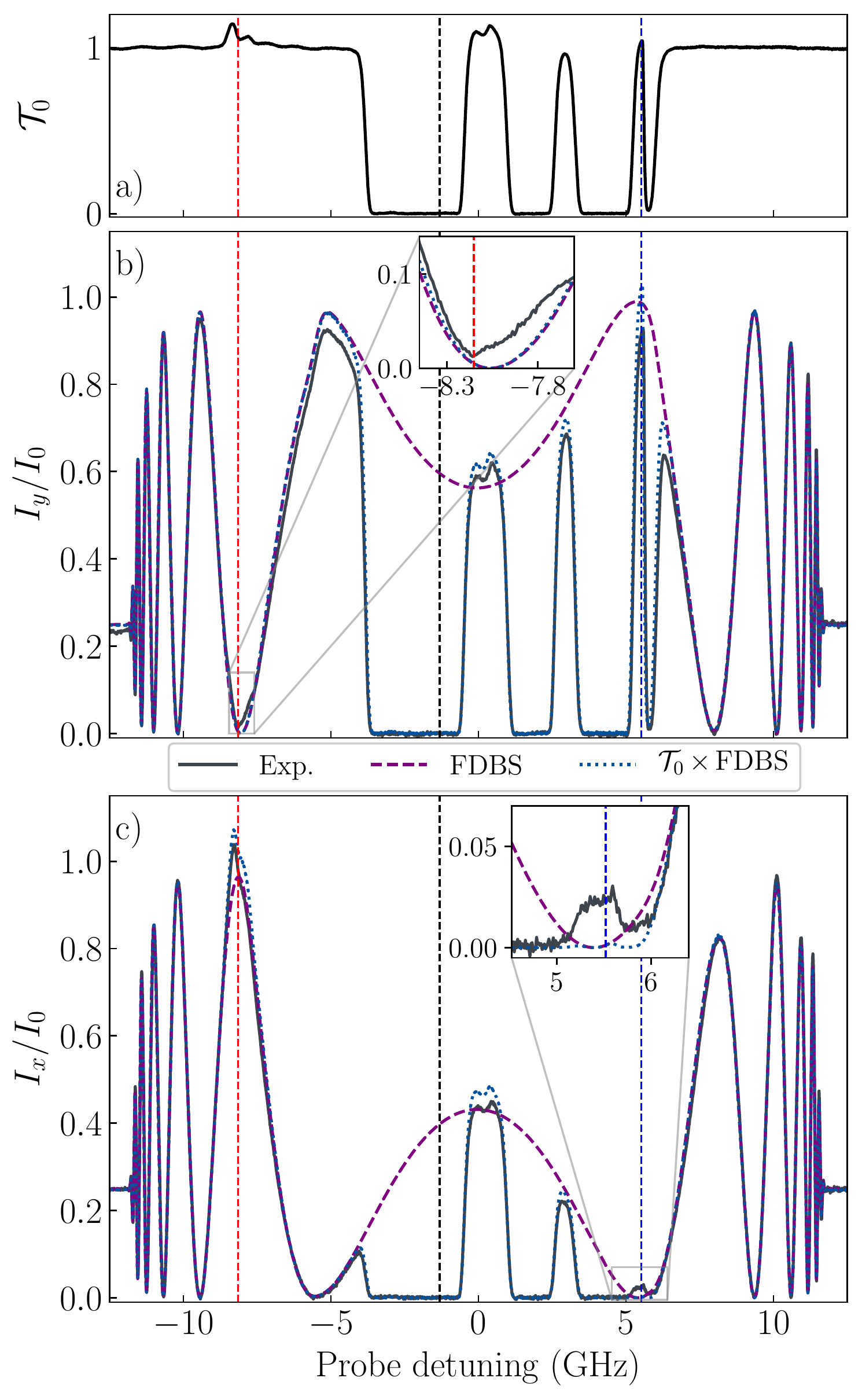}
	\caption{a) Experimental transmission spectrum ($\mathcal{T}_0$) of the probe laser through both the $^{87}$Rb hollow-core fiber cell \mbox{(HC-FC)} and the $^{85}$Rb pump filter cell. Panels b) and c) show spectra in units of the initial intensity $I_0$ polarized in the $y$- and $x$-directions respectively. The black curves show the experimental spectra after the probe laser has traversed all three cells. The dashed purple curves show the response of the Faraday dichroic beam splitter (FDBS) only. The blue dotted curve shows the measurement expected if all the light detected were at the same frequency as the probe detuning ($\mathcal{T}_0\times$FDBS). The vertical dashed lines show the frequencies $\nu_1$, $\nu_p$ and $\nu_2$ according to figure~\ref{fig:leveldiagram}. The insets in panels b) and c) highlight regions where a clear excess from the $\mathcal{T}_0\times$FDBS curves is measured.}
	\label{fig:InterfacedSpectra}
\end{figure}

Panel a) of figure~\ref{fig:InterfacedSpectra} shows the measured transmission spectrum after the first two vapour cells ($\mathcal{T}_0$) but before the FDBS. Gain features are seen when the probe is near the two-photon resonance frequencies, $\nu_1$ and $\nu_2$. Near $\nu_1$ a peak transmission of 1.15 is seen. Previous similar experiments using a bulk vapour cells with lengths approximately 10 times shorter, show gain at a similar order of magnitude with hundreds of mW of pump power~\cite{Pooser2009}. This is in line with the length squared enhancement in efficiency expected~\cite{Bratfalean1996}, however the relatively small mode due to the fiber-core size means we are working at much higher intensities. The gain peak around $\nu_1$ exhibits Autler-Townes splitting and a light shift~\cite{Cohen-Tannoudji1996} of approximately 200\,MHz of the strongest peak. Competition between FWM gain and Raman absorption near $\nu_2$ means the gain peak is somewhat suppressed and the spectrum is more complicated due to these additional features~\cite{McCormick2006}. Both the gain features around the two-photon resonances show significant power broadening, but nevertheless, the relatively broadband Faraday rotation features of the FDBS capture much of the light across the gain bandwidths.

The gain features observed can be the result of FWM or Raman processes. Using the FDBS we can distinguish between the two processes. This is because Raman processes do no produce observable light at any other frequencies other than that of the probe and therefore the overall spectral response after should simply be the product of the spectra from individual cells, i.e. $I_{x,y} = I_{x,y}^{\mathrm{FDBS}}\mathcal{T}_0$, where $I_{x,y}^{\mathrm{FDBS}}$ denotes the intensity spectra for the FDBS device only. Any measured deviation from this simple product is evidence for light having been produced at another frequency. Panels b) and c) of figure~\ref{fig:InterfacedSpectra} show the $I_y$ and $I_x$ outputs after the probe has traversed the FDBS. Deviations are indeed seen, however, due to the laser power changing throughout the scan there is a residual error of up to 5\,\% of our estimate of the initial intensity spectrum $I_0$. This means that deviations across some parts of the spectrum are simply measurement artefacts. However, the effect of this measurement error is almost entirely suppressed around the regions where the spectra are near zero. The insets in panels b) and c) of figure~\ref{fig:InterfacedSpectra} shows an expanded view around the two-photon resonances for the outputs where almost no light should be seen for those probe detuning values. A clear excess of approximately 2\,\% is seen in both cases, which indicates light having been generated from FWM. From these results we can estimate that 20\,\% of the gain observed is due to FWM while 80\,\% is from Raman scattering. For a heralded single photon source, this level of Raman scattering would significantly degrade the performance since it is a source of photons in the two channels which are uncorrelated. Further work is necessary to ascertain the reason for the low proportion of photon pairs from FWM compared to Raman scattering, but it is likely to be due to a loss of ground state coherence from the atoms colliding with the fiber walls, despite the fact that the thin walls of the kagome structured fibers have shown some anti-relaxation properties~\cite{Version2019}. A solution could be to introduce a buffer gas to the fiber cells, which should dramatically increase the ground state coherence~\cite{Brandt1997}.

\section{Conclusions and outlook}

We have presented a Faraday dichroic beam splitter designed to separate signal and idler fields from a four-wave mixing process in $^{87}$Rb. The device exploits the Faraday effect in a rubidium atomic vapour to rotate the plane of polarization of one frequency $90^\circ$ more than another. Using computational optimization with a tailored cost function, we have found the optimal magnetic field and temperature of the vapour cell to give excellent transmission and separation of the two frequencies. The device has been tested experimentally with the atomic vapour found to transmit at 97 and 99\,\% of light for the lower and higher frequency channel respectively, with corresponding extinction ratios of $(-26.3\pm0.1)$ and $(-21.2\pm0.1)$\,dB. Furthermore, we applied the Faraday dichroic beam splitter in an envisaged use case, namely that of separating signal and idler fields from four-wave mixing in a $^{87}$Rb medium loaded into a hollow-core fiber, which precludes using spatial separation from phase matching to separate the two fields. This allowed the estimation of the amount of four-wave mixing occurring in relative to that Raman scattering, which we found at a ratio of 1:5. We expect that if the Raman scattering can be suppressed, the Faraday dichroic beam splitter could be used to observe heralded single photons from the $^{87}$Rb loaded hollow-core fiber system.

\begin{acknowledgments}
The authors thank F. Schreiber for technical support in constructing the fiber cells, and N. Y. Joly and P. St. J. Russell of the Max Planck Institute for the Science of Light (Erlangen) for providing the hollow-core photonic crystal fiber. This project has received funding from the European Union’s Horizon 2020 research and innovation programme under the Marie Sklodowska-Curie grant agreement No 789642.
\end{acknowledgments}

\bibliography{main}

\begin{thebibliography}{50}%
\makeatletter
\providecommand \@ifxundefined [1]{%
 \@ifx{#1\undefined}
}%
\providecommand \@ifnum [1]{%
 \ifnum #1\expandafter \@firstoftwo
 \else \expandafter \@secondoftwo
 \fi
}%
\providecommand \@ifx [1]{%
 \ifx #1\expandafter \@firstoftwo
 \else \expandafter \@secondoftwo
 \fi
}%
\providecommand \natexlab [1]{#1}%
\providecommand \enquote  [1]{``#1''}%
\providecommand \bibnamefont  [1]{#1}%
\providecommand \bibfnamefont [1]{#1}%
\providecommand \citenamefont [1]{#1}%
\providecommand \href@noop [0]{\@secondoftwo}%
\providecommand \href [0]{\begingroup \@sanitize@url \@href}%
\providecommand \@href[1]{\@@startlink{#1}\@@href}%
\providecommand \@@href[1]{\endgroup#1\@@endlink}%
\providecommand \@sanitize@url [0]{\catcode `\\12\catcode `\$12\catcode
  `\&12\catcode `\#12\catcode `\^12\catcode `\_12\catcode `\%12\relax}%
\providecommand \@@startlink[1]{}%
\providecommand \@@endlink[0]{}%
\providecommand \url  [0]{\begingroup\@sanitize@url \@url }%
\providecommand \@url [1]{\endgroup\@href {#1}{\urlprefix }}%
\providecommand \urlprefix  [0]{URL }%
\providecommand \Eprint [0]{\href }%
\providecommand \doibase [0]{https://doi.org/}%
\providecommand \selectlanguage [0]{\@gobble}%
\providecommand \bibinfo  [0]{\@secondoftwo}%
\providecommand \bibfield  [0]{\@secondoftwo}%
\providecommand \translation [1]{[#1]}%
\providecommand \BibitemOpen [0]{}%
\providecommand \bibitemStop [0]{}%
\providecommand \bibitemNoStop [0]{.\EOS\space}%
\providecommand \EOS [0]{\spacefactor3000\relax}%
\providecommand \BibitemShut  [1]{\csname bibitem#1\endcsname}%
\let\auto@bib@innerbib\@empty
\bibitem [{\citenamefont {Benabid}\ \emph {et~al.}(2005)\citenamefont
  {Benabid}, \citenamefont {Couny}, \citenamefont {Knight}, \citenamefont
  {Birks},\ and\ \citenamefont {Russell}}]{Benabid2005}%
  \BibitemOpen
  \bibfield  {author} {\bibinfo {author} {\bibfnamefont {F.}~\bibnamefont
  {Benabid}}, \bibinfo {author} {\bibfnamefont {F.}~\bibnamefont {Couny}},
  \bibinfo {author} {\bibfnamefont {J.~C.}\ \bibnamefont {Knight}}, \bibinfo
  {author} {\bibfnamefont {T.~A.}\ \bibnamefont {Birks}},\ and\ \bibinfo
  {author} {\bibfnamefont {P.~{\relax St}.~J.}\ \bibnamefont {Russell}},\
  }\bibfield  {title} {\bibinfo {title} {{Compact, stable and efficient
  all-fibre gas cells using hollow-core photonic crystal fibres}},\ }\href
  {https://doi.org/10.1038/nature03349} {\bibfield  {journal} {\bibinfo
  {journal} {Nature}\ }\textbf {\bibinfo {volume} {434}},\ \bibinfo {pages}
  {488} (\bibinfo {year} {2005})}\BibitemShut {NoStop}%
\bibitem [{\citenamefont {Russell}(2003)}]{Fevrier2011}%
  \BibitemOpen
  \bibfield  {author} {\bibinfo {author} {\bibfnamefont {P.}~\bibnamefont
  {Russell}},\ }\bibfield  {title} {\bibinfo {title} {{Photonic Crystal
  Fibers}},\ }\href {https://doi.org/10.1126/science.1079280} {\bibfield
  {journal} {\bibinfo  {journal} {Science}\ }\textbf {\bibinfo {volume}
  {299}},\ \bibinfo {pages} {358} (\bibinfo {year} {2003})}\BibitemShut
  {NoStop}%
\bibitem [{\citenamefont {Debord}\ \emph {et~al.}(2019)\citenamefont {Debord},
  \citenamefont {Amrani}, \citenamefont {Vincetti}, \citenamefont
  {G{\'{e}}r{\^{o}}me},\ and\ \citenamefont {Benabid}}]{Debord2019}%
  \BibitemOpen
  \bibfield  {author} {\bibinfo {author} {\bibfnamefont {B.}~\bibnamefont
  {Debord}}, \bibinfo {author} {\bibfnamefont {F.}~\bibnamefont {Amrani}},
  \bibinfo {author} {\bibfnamefont {L.}~\bibnamefont {Vincetti}}, \bibinfo
  {author} {\bibfnamefont {F.}~\bibnamefont {G{\'{e}}r{\^{o}}me}},\ and\
  \bibinfo {author} {\bibfnamefont {F.}~\bibnamefont {Benabid}},\ }\bibfield
  {title} {\bibinfo {title} {{Hollow-Core Fiber Technology: The Rising of
  “Gas Photonics”}},\ }\href {https://doi.org/10.3390/fib7020016}
  {\bibfield  {journal} {\bibinfo  {journal} {Fibers}\ }\textbf {\bibinfo
  {volume} {7}},\ \bibinfo {pages} {16} (\bibinfo {year} {2019})}\BibitemShut
  {NoStop}%
\bibitem [{\citenamefont {Horan}\ \emph {et~al.}(2012)\citenamefont {Horan},
  \citenamefont {Ruth},\ and\ \citenamefont {Garcia~Gunning}}]{Horan2012}%
  \BibitemOpen
  \bibfield  {author} {\bibinfo {author} {\bibfnamefont {L.~E.}\ \bibnamefont
  {Horan}}, \bibinfo {author} {\bibfnamefont {A.~A.}\ \bibnamefont {Ruth}},\
  and\ \bibinfo {author} {\bibfnamefont {F.~C.}\ \bibnamefont
  {Garcia~Gunning}},\ }\bibfield  {title} {\bibinfo {title} {{Hollow core
  photonic crystal fiber based viscometer with Raman spectroscopy}},\ }\href
  {https://doi.org/10.1063/1.4771659} {\bibfield  {journal} {\bibinfo
  {journal} {J. Chem. Phys.}\ }\textbf {\bibinfo {volume} {137}},\ \bibinfo
  {pages} {224504} (\bibinfo {year} {2012})}\BibitemShut {NoStop}%
\bibitem [{\citenamefont {Ghosh}\ \emph {et~al.}(2006)\citenamefont {Ghosh},
  \citenamefont {Bhagwat}, \citenamefont {Kyle~Renshaw}, \citenamefont {Goh},
  \citenamefont {Gaeta},\ and\ \citenamefont {Kirby}}]{Ghosh2006a}%
  \BibitemOpen
  \bibfield  {author} {\bibinfo {author} {\bibfnamefont {S.}~\bibnamefont
  {Ghosh}}, \bibinfo {author} {\bibfnamefont {A.~R.}\ \bibnamefont {Bhagwat}},
  \bibinfo {author} {\bibfnamefont {C.}~\bibnamefont {Kyle~Renshaw}}, \bibinfo
  {author} {\bibfnamefont {S.}~\bibnamefont {Goh}}, \bibinfo {author}
  {\bibfnamefont {A.~L.}\ \bibnamefont {Gaeta}},\ and\ \bibinfo {author}
  {\bibfnamefont {B.~J.}\ \bibnamefont {Kirby}},\ }\bibfield  {title} {\bibinfo
  {title} {{Low-Light-Level Optical Interactions with Rubidium Vapor in a
  Photonic Band-Gap Fiber}},\ }\href
  {https://doi.org/10.1103/PhysRevLett.97.023603} {\bibfield  {journal}
  {\bibinfo  {journal} {Phys. Rev. Lett.}\ }\textbf {\bibinfo {volume} {97}},\
  \bibinfo {pages} {023603} (\bibinfo {year} {2006})}\BibitemShut {NoStop}%
\bibitem [{\citenamefont {Londero}\ \emph {et~al.}(2009)\citenamefont
  {Londero}, \citenamefont {Venkataraman}, \citenamefont {Bhagwat},
  \citenamefont {Slepkov},\ and\ \citenamefont {Gaeta}}]{Londero2009}%
  \BibitemOpen
  \bibfield  {author} {\bibinfo {author} {\bibfnamefont {P.}~\bibnamefont
  {Londero}}, \bibinfo {author} {\bibfnamefont {V.}~\bibnamefont
  {Venkataraman}}, \bibinfo {author} {\bibfnamefont {A.~R.}\ \bibnamefont
  {Bhagwat}}, \bibinfo {author} {\bibfnamefont {A.~D.}\ \bibnamefont
  {Slepkov}},\ and\ \bibinfo {author} {\bibfnamefont {A.~L.}\ \bibnamefont
  {Gaeta}},\ }\bibfield  {title} {\bibinfo {title} {{Ultralow-Power Four-Wave
  Mixing with Rb in a Hollow-Core Photonic Band-Gap Fiber}},\ }\href
  {https://doi.org/10.1103/PhysRevLett.103.043602} {\bibfield  {journal}
  {\bibinfo  {journal} {Phys. Rev. Lett.}\ }\textbf {\bibinfo {volume} {103}},\
  \bibinfo {pages} {043602} (\bibinfo {year} {2009})}\BibitemShut {NoStop}%
\bibitem [{\citenamefont {Perrella}\ \emph {et~al.}(2013)\citenamefont
  {Perrella}, \citenamefont {Light}, \citenamefont {Anstie}, \citenamefont
  {Benabid}, \citenamefont {Stace}, \citenamefont {White},\ and\ \citenamefont
  {Luiten}}]{Perrella2013a}%
  \BibitemOpen
  \bibfield  {author} {\bibinfo {author} {\bibfnamefont {C.}~\bibnamefont
  {Perrella}}, \bibinfo {author} {\bibfnamefont {P.~S.}\ \bibnamefont {Light}},
  \bibinfo {author} {\bibfnamefont {J.~D.}\ \bibnamefont {Anstie}}, \bibinfo
  {author} {\bibfnamefont {F.}~\bibnamefont {Benabid}}, \bibinfo {author}
  {\bibfnamefont {T.~M.}\ \bibnamefont {Stace}}, \bibinfo {author}
  {\bibfnamefont {A.~G.}\ \bibnamefont {White}},\ and\ \bibinfo {author}
  {\bibfnamefont {A.~N.}\ \bibnamefont {Luiten}},\ }\bibfield  {title}
  {\bibinfo {title} {{High-efficiency cross-phase modulation in a gas-filled
  waveguide}},\ }\href {https://doi.org/10.1103/PhysRevA.88.013819} {\bibfield
  {journal} {\bibinfo  {journal} {Phys. Rev. A}\ }\textbf {\bibinfo {volume}
  {88}},\ \bibinfo {pages} {013819} (\bibinfo {year} {2013})}\BibitemShut
  {NoStop}%
\bibitem [{\citenamefont {Sprague}\ \emph {et~al.}(2014)\citenamefont
  {Sprague}, \citenamefont {Michelberger}, \citenamefont {Champion},
  \citenamefont {England}, \citenamefont {Nunn}, \citenamefont {Jin},
  \citenamefont {Kolthammer}, \citenamefont {Abdolvand}, \citenamefont
  {Russell},\ and\ \citenamefont {Walmsley}}]{Sprague2014}%
  \BibitemOpen
  \bibfield  {author} {\bibinfo {author} {\bibfnamefont {M.~R.}\ \bibnamefont
  {Sprague}}, \bibinfo {author} {\bibfnamefont {P.~S.}\ \bibnamefont
  {Michelberger}}, \bibinfo {author} {\bibfnamefont {T.~F.~M.}\ \bibnamefont
  {Champion}}, \bibinfo {author} {\bibfnamefont {D.~G.}\ \bibnamefont
  {England}}, \bibinfo {author} {\bibfnamefont {J.}~\bibnamefont {Nunn}},
  \bibinfo {author} {\bibfnamefont {X.-M.}\ \bibnamefont {Jin}}, \bibinfo
  {author} {\bibfnamefont {W.~S.}\ \bibnamefont {Kolthammer}}, \bibinfo
  {author} {\bibfnamefont {A.}~\bibnamefont {Abdolvand}}, \bibinfo {author}
  {\bibfnamefont {P.~{\relax St}.~J.}\ \bibnamefont {Russell}},\ and\ \bibinfo
  {author} {\bibfnamefont {I.~A.}\ \bibnamefont {Walmsley}},\ }\bibfield
  {title} {\bibinfo {title} {{Broadband single-photon-level memory in a
  hollow-core photonic crystal fibre}},\ }\href
  {https://doi.org/10.1038/nphoton.2014.45} {\bibfield  {journal} {\bibinfo
  {journal} {Nat. Photonics}\ }\textbf {\bibinfo {volume} {8}},\ \bibinfo
  {pages} {287} (\bibinfo {year} {2014})}\BibitemShut {NoStop}%
\bibitem [{\citenamefont {Cordier}\ \emph {et~al.}(2020)\citenamefont
  {Cordier}, \citenamefont {Delaye}, \citenamefont {G{\'{e}}r{\^{o}}me},
  \citenamefont {Benabid},\ and\ \citenamefont {Zaquine}}]{Cordier2020}%
  \BibitemOpen
  \bibfield  {author} {\bibinfo {author} {\bibfnamefont {M.}~\bibnamefont
  {Cordier}}, \bibinfo {author} {\bibfnamefont {P.}~\bibnamefont {Delaye}},
  \bibinfo {author} {\bibfnamefont {F.}~\bibnamefont {G{\'{e}}r{\^{o}}me}},
  \bibinfo {author} {\bibfnamefont {F.}~\bibnamefont {Benabid}},\ and\ \bibinfo
  {author} {\bibfnamefont {I.}~\bibnamefont {Zaquine}},\ }\bibfield  {title}
  {\bibinfo {title} {{Raman-free fibered photon-pair source}},\ }\href
  {https://doi.org/10.1038/s41598-020-58229-7} {\bibfield  {journal} {\bibinfo
  {journal} {Sci. Rep.}\ }\textbf {\bibinfo {volume} {10}},\ \bibinfo {pages}
  {1650} (\bibinfo {year} {2020})}\BibitemShut {NoStop}%
\bibitem [{\citenamefont {McCormick}\ \emph {et~al.}(2007)\citenamefont
  {McCormick}, \citenamefont {Boyer}, \citenamefont {Arimondo},\ and\
  \citenamefont {Lett}}]{McCormick2006}%
  \BibitemOpen
  \bibfield  {author} {\bibinfo {author} {\bibfnamefont {C.~F.}\ \bibnamefont
  {McCormick}}, \bibinfo {author} {\bibfnamefont {V.}~\bibnamefont {Boyer}},
  \bibinfo {author} {\bibfnamefont {E.}~\bibnamefont {Arimondo}},\ and\
  \bibinfo {author} {\bibfnamefont {P.~D.}\ \bibnamefont {Lett}},\ }\bibfield
  {title} {\bibinfo {title} {{Strong relative intensity squeezing by four-wave
  mixing in rubidium vapor}},\ }\href {https://doi.org/10.1364/OL.32.000178}
  {\bibfield  {journal} {\bibinfo  {journal} {Opt. Lett.}\ }\textbf {\bibinfo
  {volume} {32}},\ \bibinfo {pages} {178} (\bibinfo {year} {2007})}\BibitemShut
  {NoStop}%
\bibitem [{\citenamefont {MacRae}\ \emph {et~al.}(2012)\citenamefont {MacRae},
  \citenamefont {Brannan}, \citenamefont {Achal},\ and\ \citenamefont
  {Lvovsky}}]{MacRae2012}%
  \BibitemOpen
  \bibfield  {author} {\bibinfo {author} {\bibfnamefont {A.}~\bibnamefont
  {MacRae}}, \bibinfo {author} {\bibfnamefont {T.}~\bibnamefont {Brannan}},
  \bibinfo {author} {\bibfnamefont {R.}~\bibnamefont {Achal}},\ and\ \bibinfo
  {author} {\bibfnamefont {A.~I.}\ \bibnamefont {Lvovsky}},\ }\bibfield
  {title} {\bibinfo {title} {{Tomography of a High-Purity Narrowband Photon
  from a Transient Atomic Collective Excitation}},\ }\href
  {https://doi.org/10.1103/PhysRevLett.109.033601} {\bibfield  {journal}
  {\bibinfo  {journal} {Phys. Rev. Lett.}\ }\textbf {\bibinfo {volume} {109}},\
  \bibinfo {pages} {033601} (\bibinfo {year} {2012})}\BibitemShut {NoStop}%
\bibitem [{\citenamefont {Shu}\ \emph {et~al.}(2016)\citenamefont {Shu},
  \citenamefont {Chen}, \citenamefont {Chow}, \citenamefont {Zhu},
  \citenamefont {Xiao}, \citenamefont {Loy},\ and\ \citenamefont
  {Du}}]{Shu2016}%
  \BibitemOpen
  \bibfield  {author} {\bibinfo {author} {\bibfnamefont {C.}~\bibnamefont
  {Shu}}, \bibinfo {author} {\bibfnamefont {P.}~\bibnamefont {Chen}}, \bibinfo
  {author} {\bibfnamefont {T.~K.~A.}\ \bibnamefont {Chow}}, \bibinfo {author}
  {\bibfnamefont {L.}~\bibnamefont {Zhu}}, \bibinfo {author} {\bibfnamefont
  {Y.}~\bibnamefont {Xiao}}, \bibinfo {author} {\bibfnamefont {M.~M.~T.}\
  \bibnamefont {Loy}},\ and\ \bibinfo {author} {\bibfnamefont {S.}~\bibnamefont
  {Du}},\ }\bibfield  {title} {\bibinfo {title} {{Subnatural-linewidth
  biphotons from a Doppler-broadened hot atomic vapour cell}},\ }\href
  {https://doi.org/10.1038/ncomms12783} {\bibfield  {journal} {\bibinfo
  {journal} {Nat. Commun.}\ }\textbf {\bibinfo {volume} {7}},\ \bibinfo {pages}
  {12783} (\bibinfo {year} {2016})}\BibitemShut {NoStop}%
\bibitem [{\citenamefont {Whiting}\ \emph {et~al.}(2017)\citenamefont
  {Whiting}, \citenamefont {{\v{S}}ibali{\'{c}}}, \citenamefont {Keaveney},
  \citenamefont {Adams},\ and\ \citenamefont {Hughes}}]{Whiting2017}%
  \BibitemOpen
  \bibfield  {author} {\bibinfo {author} {\bibfnamefont {D.~J.}\ \bibnamefont
  {Whiting}}, \bibinfo {author} {\bibfnamefont {N.}~\bibnamefont
  {{\v{S}}ibali{\'{c}}}}, \bibinfo {author} {\bibfnamefont {J.}~\bibnamefont
  {Keaveney}}, \bibinfo {author} {\bibfnamefont {C.~S.}\ \bibnamefont
  {Adams}},\ and\ \bibinfo {author} {\bibfnamefont {I.~G.}\ \bibnamefont
  {Hughes}},\ }\bibfield  {title} {\bibinfo {title} {{Single-Photon
  Interference due to Motion in an Atomic Collective Excitation}},\ }\href
  {https://doi.org/10.1103/PhysRevLett.118.253601} {\bibfield  {journal}
  {\bibinfo  {journal} {Phys. Rev. Lett.}\ }\textbf {\bibinfo {volume} {118}},\
  \bibinfo {pages} {253601} (\bibinfo {year} {2017})}\BibitemShut {NoStop}%
\bibitem [{\citenamefont {Ripka}\ \emph {et~al.}(2018)\citenamefont {Ripka},
  \citenamefont {K{\"{u}}bler}, \citenamefont {L{\"{o}}w},\ and\ \citenamefont
  {Pfau}}]{Ripka2018}%
  \BibitemOpen
  \bibfield  {author} {\bibinfo {author} {\bibfnamefont {F.}~\bibnamefont
  {Ripka}}, \bibinfo {author} {\bibfnamefont {H.}~\bibnamefont {K{\"{u}}bler}},
  \bibinfo {author} {\bibfnamefont {R.}~\bibnamefont {L{\"{o}}w}},\ and\
  \bibinfo {author} {\bibfnamefont {T.}~\bibnamefont {Pfau}},\ }\bibfield
  {title} {\bibinfo {title} {{A room-temperature single-photon source based on
  strongly interacting Rydberg atoms}},\ }\href
  {https://doi.org/10.1126/science.aau1949} {\bibfield  {journal} {\bibinfo
  {journal} {Science}\ }\textbf {\bibinfo {volume} {362}},\ \bibinfo {pages}
  {446} (\bibinfo {year} {2018})}\BibitemShut {NoStop}%
\bibitem [{\citenamefont {Julsgaard}\ \emph {et~al.}(2004)\citenamefont
  {Julsgaard}, \citenamefont {Sherson}, \citenamefont {Ignacio~Cirac},
  \citenamefont {Fiur{\'{a}}{\v{s}}ek},\ and\ \citenamefont
  {Polzik}}]{Julsgaard2004}%
  \BibitemOpen
  \bibfield  {author} {\bibinfo {author} {\bibfnamefont {B.}~\bibnamefont
  {Julsgaard}}, \bibinfo {author} {\bibfnamefont {J.}~\bibnamefont {Sherson}},
  \bibinfo {author} {\bibfnamefont {J.}~\bibnamefont {Ignacio~Cirac}}, \bibinfo
  {author} {\bibfnamefont {J.}~\bibnamefont {Fiur{\'{a}}{\v{s}}ek}},\ and\
  \bibinfo {author} {\bibfnamefont {E.~S.}\ \bibnamefont {Polzik}},\ }\bibfield
   {title} {\bibinfo {title} {{Experimental demonstration of quantum memory for
  light.}},\ }\href {https://doi.org/10.1038/nature03064} {\bibfield  {journal}
  {\bibinfo  {journal} {Nature}\ }\textbf {\bibinfo {volume} {432}},\ \bibinfo
  {pages} {482} (\bibinfo {year} {2004})}\BibitemShut {NoStop}%
\bibitem [{\citenamefont {Hosseini}\ \emph {et~al.}(2011)\citenamefont
  {Hosseini}, \citenamefont {Sparkes}, \citenamefont {Campbell}, \citenamefont
  {Lam},\ and\ \citenamefont {Buchler}}]{Hosseini2011}%
  \BibitemOpen
  \bibfield  {author} {\bibinfo {author} {\bibfnamefont {M.}~\bibnamefont
  {Hosseini}}, \bibinfo {author} {\bibfnamefont {B.~M.}\ \bibnamefont
  {Sparkes}}, \bibinfo {author} {\bibfnamefont {G.}~\bibnamefont {Campbell}},
  \bibinfo {author} {\bibfnamefont {P.~K.}\ \bibnamefont {Lam}},\ and\ \bibinfo
  {author} {\bibfnamefont {B.~C.}\ \bibnamefont {Buchler}},\ }\bibfield
  {title} {\bibinfo {title} {{High efficiency coherent optical memory with warm
  rubidium vapour}},\ }\href {https://doi.org/10.1038/ncomms1175} {\bibfield
  {journal} {\bibinfo  {journal} {Nat. Commun.}\ }\textbf {\bibinfo {volume}
  {2}},\ \bibinfo {pages} {174} (\bibinfo {year} {2011})}\BibitemShut {NoStop}%
\bibitem [{\citenamefont {Michelberger}\ \emph {et~al.}(2015)\citenamefont
  {Michelberger}, \citenamefont {Champion}, \citenamefont {Sprague},
  \citenamefont {Kaczmarek}, \citenamefont {Barbieri}, \citenamefont {Jin},
  \citenamefont {England}, \citenamefont {Kolthammer}, \citenamefont
  {Saunders}, \citenamefont {Nunn},\ and\ \citenamefont
  {Walmsley}}]{Michelberger2015a}%
  \BibitemOpen
  \bibfield  {author} {\bibinfo {author} {\bibfnamefont {P.~S.}\ \bibnamefont
  {Michelberger}}, \bibinfo {author} {\bibfnamefont {T.~F.~M.}\ \bibnamefont
  {Champion}}, \bibinfo {author} {\bibfnamefont {M.~R.}\ \bibnamefont
  {Sprague}}, \bibinfo {author} {\bibfnamefont {K.~T.}\ \bibnamefont
  {Kaczmarek}}, \bibinfo {author} {\bibfnamefont {M.}~\bibnamefont {Barbieri}},
  \bibinfo {author} {\bibfnamefont {X.~M.}\ \bibnamefont {Jin}}, \bibinfo
  {author} {\bibfnamefont {D.~G.}\ \bibnamefont {England}}, \bibinfo {author}
  {\bibfnamefont {W.~S.}\ \bibnamefont {Kolthammer}}, \bibinfo {author}
  {\bibfnamefont {D.~J.}\ \bibnamefont {Saunders}}, \bibinfo {author}
  {\bibfnamefont {J.}~\bibnamefont {Nunn}},\ and\ \bibinfo {author}
  {\bibfnamefont {I.~A.}\ \bibnamefont {Walmsley}},\ }\bibfield  {title}
  {\bibinfo {title} {{Interfacing GHz-bandwidth heralded single photons with a
  warm vapour Raman memory}},\ }\href
  {https://doi.org/10.1088/1367-2630/17/4/043006} {\bibfield  {journal}
  {\bibinfo  {journal} {New J. Phys.}\ }\textbf {\bibinfo {volume} {17}},\
  \bibinfo {pages} {043006} (\bibinfo {year} {2015})}\BibitemShut {NoStop}%
\bibitem [{\citenamefont {Podhora}\ \emph {et~al.}(2017)\citenamefont
  {Podhora}, \citenamefont {Ob{\v{s}}il}, \citenamefont {Straka}, \citenamefont
  {Je{\v{z}}ek},\ and\ \citenamefont {Slodi{\v{c}}ka}}]{Podhora2017}%
  \BibitemOpen
  \bibfield  {author} {\bibinfo {author} {\bibfnamefont {L.}~\bibnamefont
  {Podhora}}, \bibinfo {author} {\bibfnamefont {P.}~\bibnamefont
  {Ob{\v{s}}il}}, \bibinfo {author} {\bibfnamefont {I.}~\bibnamefont {Straka}},
  \bibinfo {author} {\bibfnamefont {M.}~\bibnamefont {Je{\v{z}}ek}},\ and\
  \bibinfo {author} {\bibfnamefont {L.}~\bibnamefont {Slodi{\v{c}}ka}},\
  }\bibfield  {title} {\bibinfo {title} {{Nonclassical photon pairs from warm
  atomic vapor using a single driving laser}},\ }\href
  {https://doi.org/10.1364/OE.25.031230} {\bibfield  {journal} {\bibinfo
  {journal} {Opt. Express}\ }\textbf {\bibinfo {volume} {25}},\ \bibinfo
  {pages} {31230} (\bibinfo {year} {2017})}\BibitemShut {NoStop}%
\bibitem [{\citenamefont {Bratfalean}\ and\ \citenamefont
  {Ewart}(1996)}]{Bratfalean1996}%
  \BibitemOpen
  \bibfield  {author} {\bibinfo {author} {\bibfnamefont {R.}~\bibnamefont
  {Bratfalean}}\ and\ \bibinfo {author} {\bibfnamefont {P.}~\bibnamefont
  {Ewart}},\ }\bibfield  {title} {\bibinfo {title} {{The dependence of
  broadband four-wave mixing signal intensity on the length of the interaction
  region}},\ }\href {https://doi.org/10.1080/09500349608230678} {\bibfield
  {journal} {\bibinfo  {journal} {J. Mod. Opt.}\ }\textbf {\bibinfo {volume}
  {43}},\ \bibinfo {pages} {2523} (\bibinfo {year} {1996})}\BibitemShut
  {NoStop}%
\bibitem [{\citenamefont {Palittapongarnpim}\ \emph {et~al.}(2012)\citenamefont
  {Palittapongarnpim}, \citenamefont {MacRae},\ and\ \citenamefont
  {Lvovsky}}]{Palittapongarnpim2012}%
  \BibitemOpen
  \bibfield  {author} {\bibinfo {author} {\bibfnamefont {P.}~\bibnamefont
  {Palittapongarnpim}}, \bibinfo {author} {\bibfnamefont {A.}~\bibnamefont
  {MacRae}},\ and\ \bibinfo {author} {\bibfnamefont {A.~I.}\ \bibnamefont
  {Lvovsky}},\ }\bibfield  {title} {\bibinfo {title} {{Note: a monolithic
  filter cavity for experiments in quantum optics}},\ }\href
  {https://doi.org/10.1063/1.4726458} {\bibfield  {journal} {\bibinfo
  {journal} {Rev. Sci. Instrum.}\ }\textbf {\bibinfo {volume} {83}},\ \bibinfo
  {pages} {066101} (\bibinfo {year} {2012})}\BibitemShut {NoStop}%
\bibitem [{\citenamefont {Weller}\ \emph
  {et~al.}(2012{\natexlab{a}})\citenamefont {Weller}, \citenamefont
  {Kleinbach}, \citenamefont {Zentile}, \citenamefont {Knappe}, \citenamefont
  {Hughes},\ and\ \citenamefont {Adams}}]{Weller2012d}%
  \BibitemOpen
  \bibfield  {author} {\bibinfo {author} {\bibfnamefont {L.}~\bibnamefont
  {Weller}}, \bibinfo {author} {\bibfnamefont {K.~S.}\ \bibnamefont
  {Kleinbach}}, \bibinfo {author} {\bibfnamefont {M.~A.}\ \bibnamefont
  {Zentile}}, \bibinfo {author} {\bibfnamefont {S.}~\bibnamefont {Knappe}},
  \bibinfo {author} {\bibfnamefont {I.~G.}\ \bibnamefont {Hughes}},\ and\
  \bibinfo {author} {\bibfnamefont {C.~S.}\ \bibnamefont {Adams}},\ }\bibfield
  {title} {\bibinfo {title} {{Optical isolator using an atomic vapor in the
  hyperfine Paschen–Back regime}},\ }\href
  {https://doi.org/10.1364/OL.37.003405} {\bibfield  {journal} {\bibinfo
  {journal} {Opt. Lett.}\ }\textbf {\bibinfo {volume} {37}},\ \bibinfo {pages}
  {3405} (\bibinfo {year} {2012}{\natexlab{a}})}\BibitemShut {NoStop}%
\bibitem [{\citenamefont {Gerhardt}(2018)}]{Gerhardt2018}%
  \BibitemOpen
  \bibfield  {author} {\bibinfo {author} {\bibfnamefont {I.}~\bibnamefont
  {Gerhardt}},\ }\bibfield  {title} {\bibinfo {title} {{How anomalous is my
  Faraday filter?}},\ }\href {https://doi.org/10.1364/OL.43.005295} {\bibfield
  {journal} {\bibinfo  {journal} {Opt. Lett.}\ }\textbf {\bibinfo {volume}
  {43}},\ \bibinfo {pages} {5295} (\bibinfo {year} {2018})}\BibitemShut
  {NoStop}%
\bibitem [{\citenamefont {Abel}\ \emph {et~al.}(2009)\citenamefont {Abel},
  \citenamefont {Krohn}, \citenamefont {Siddons}, \citenamefont {Hughes},\ and\
  \citenamefont {Adams}}]{Abel2009}%
  \BibitemOpen
  \bibfield  {author} {\bibinfo {author} {\bibfnamefont {R.~P.}\ \bibnamefont
  {Abel}}, \bibinfo {author} {\bibfnamefont {U.}~\bibnamefont {Krohn}},
  \bibinfo {author} {\bibfnamefont {P.}~\bibnamefont {Siddons}}, \bibinfo
  {author} {\bibfnamefont {I.~G.}\ \bibnamefont {Hughes}},\ and\ \bibinfo
  {author} {\bibfnamefont {C.~S.}\ \bibnamefont {Adams}},\ }\bibfield  {title}
  {\bibinfo {title} {{Faraday dichroic beam splitter for Raman light using an
  isotopically pure alkali-metal-vapor cell}},\ }\href
  {https://doi.org/10.1364/OL.34.003071} {\bibfield  {journal} {\bibinfo
  {journal} {Opt. Lett.}\ }\textbf {\bibinfo {volume} {34}},\ \bibinfo {pages}
  {3071} (\bibinfo {year} {2009})}\BibitemShut {NoStop}%
\bibitem [{\citenamefont {Portalupi}\ \emph {et~al.}(2016)\citenamefont
  {Portalupi}, \citenamefont {Widmann}, \citenamefont {Nawrath}, \citenamefont
  {Jetter}, \citenamefont {Michler}, \citenamefont {Wrachtrup},\ and\
  \citenamefont {Gerhardt}}]{Portalupi2016}%
  \BibitemOpen
  \bibfield  {author} {\bibinfo {author} {\bibfnamefont {S.~L.}\ \bibnamefont
  {Portalupi}}, \bibinfo {author} {\bibfnamefont {M.}~\bibnamefont {Widmann}},
  \bibinfo {author} {\bibfnamefont {C.}~\bibnamefont {Nawrath}}, \bibinfo
  {author} {\bibfnamefont {M.}~\bibnamefont {Jetter}}, \bibinfo {author}
  {\bibfnamefont {P.}~\bibnamefont {Michler}}, \bibinfo {author} {\bibfnamefont
  {J.}~\bibnamefont {Wrachtrup}},\ and\ \bibinfo {author} {\bibfnamefont
  {I.}~\bibnamefont {Gerhardt}},\ }\bibfield  {title} {\bibinfo {title}
  {{Simultaneous Faraday filtering of the Mollow triplet sidebands with the
  Cs-D$_1$ clock transition}},\ }\href {https://doi.org/10.1038/ncomms13632}
  {\bibfield  {journal} {\bibinfo  {journal} {Nat. Commun.}\ }\textbf {\bibinfo
  {volume} {7}},\ \bibinfo {pages} {13632} (\bibinfo {year}
  {2016})}\BibitemShut {NoStop}%
\bibitem [{\citenamefont {Faraday}(1846)}]{Faraday1846}%
  \BibitemOpen
  \bibfield  {author} {\bibinfo {author} {\bibfnamefont {M.}~\bibnamefont
  {Faraday}},\ }\bibfield  {title} {\bibinfo {title} {{I. Experimental
  Researches in Electricity. Nineteenth Series}},\ }\href
  {https://doi.org/10.1098/rstl.1846.0001} {\bibfield  {journal} {\bibinfo
  {journal} {Philos. Trans. R. Soc. London}\ }\textbf {\bibinfo {volume}
  {136}},\ \bibinfo {pages} {1} (\bibinfo {year} {1846})}\BibitemShut {NoStop}%
\bibitem [{\citenamefont {Siddons}\ \emph {et~al.}(2009)\citenamefont
  {Siddons}, \citenamefont {Adams},\ and\ \citenamefont
  {Hughes}}]{Siddons2009}%
  \BibitemOpen
  \bibfield  {author} {\bibinfo {author} {\bibfnamefont {P.}~\bibnamefont
  {Siddons}}, \bibinfo {author} {\bibfnamefont {C.~S.}\ \bibnamefont {Adams}},\
  and\ \bibinfo {author} {\bibfnamefont {I.~G.}\ \bibnamefont {Hughes}},\
  }\bibfield  {title} {\bibinfo {title} {{Off-resonance absorption and
  dispersion in vapours of hot alkali-metal atoms}},\ }\href
  {https://doi.org/10.1088/0953-4075/42/17/175004} {\bibfield  {journal}
  {\bibinfo  {journal} {J. Phys. B: At. Mol. Opt. Phys.}\ }\textbf {\bibinfo
  {volume} {42}},\ \bibinfo {pages} {175004} (\bibinfo {year}
  {2009})}\BibitemShut {NoStop}%
\bibitem [{\citenamefont {Weller}\ \emph
  {et~al.}(2012{\natexlab{b}})\citenamefont {Weller}, \citenamefont {Dalton},
  \citenamefont {Siddons}, \citenamefont {Adams},\ and\ \citenamefont
  {Hughes}}]{Weller2012}%
  \BibitemOpen
  \bibfield  {author} {\bibinfo {author} {\bibfnamefont {L.}~\bibnamefont
  {Weller}}, \bibinfo {author} {\bibfnamefont {T.}~\bibnamefont {Dalton}},
  \bibinfo {author} {\bibfnamefont {P.}~\bibnamefont {Siddons}}, \bibinfo
  {author} {\bibfnamefont {C.~S.}\ \bibnamefont {Adams}},\ and\ \bibinfo
  {author} {\bibfnamefont {I.~G.}\ \bibnamefont {Hughes}},\ }\bibfield  {title}
  {\bibinfo {title} {{Measuring the Stokes parameters for light transmitted by
  a high-density rubidium vapour in large magnetic fields}},\ }\href
  {https://doi.org/10.1088/0953-4075/45/5/055001} {\bibfield  {journal}
  {\bibinfo  {journal} {J. Phys. B: At. Mol. Opt. Phys.}\ }\textbf {\bibinfo
  {volume} {45}},\ \bibinfo {pages} {055001} (\bibinfo {year}
  {2012}{\natexlab{b}})}\BibitemShut {NoStop}%
\bibitem [{\citenamefont {Wu}\ \emph {et~al.}(1986)\citenamefont {Wu},
  \citenamefont {Kitano}, \citenamefont {Happer}, \citenamefont {Hou},\ and\
  \citenamefont {Daniels}}]{Wu1986}%
  \BibitemOpen
  \bibfield  {author} {\bibinfo {author} {\bibfnamefont {Z.}~\bibnamefont
  {Wu}}, \bibinfo {author} {\bibfnamefont {M.}~\bibnamefont {Kitano}}, \bibinfo
  {author} {\bibfnamefont {W.}~\bibnamefont {Happer}}, \bibinfo {author}
  {\bibfnamefont {M.}~\bibnamefont {Hou}},\ and\ \bibinfo {author}
  {\bibfnamefont {J.}~\bibnamefont {Daniels}},\ }\bibfield  {title} {\bibinfo
  {title} {{Optical determination of alkali metal vapor number density using
  Faraday rotation}},\ }\href {https://doi.org/10.1364/AO.25.004483} {\bibfield
   {journal} {\bibinfo  {journal} {Appl. Opt.}\ }\textbf {\bibinfo {volume}
  {25}},\ \bibinfo {pages} {4483} (\bibinfo {year} {1986})}\BibitemShut
  {NoStop}%
\bibitem [{\citenamefont {Kemp}\ \emph {et~al.}(2011)\citenamefont {Kemp},
  \citenamefont {Hughes},\ and\ \citenamefont {Cornish}}]{Kemp2011}%
  \BibitemOpen
  \bibfield  {author} {\bibinfo {author} {\bibfnamefont {S.~L.}\ \bibnamefont
  {Kemp}}, \bibinfo {author} {\bibfnamefont {I.~G.}\ \bibnamefont {Hughes}},\
  and\ \bibinfo {author} {\bibfnamefont {S.~L.}\ \bibnamefont {Cornish}},\
  }\bibfield  {title} {\bibinfo {title} {{An analytical model of off-resonant
  Faraday rotation in hot alkali metal vapours}},\ }\href
  {https://doi.org/10.1088/0953-4075/44/23/235004} {\bibfield  {journal}
  {\bibinfo  {journal} {J. Phys. B: At. Mol. Opt. Phys.}\ }\textbf {\bibinfo
  {volume} {44}},\ \bibinfo {pages} {235004} (\bibinfo {year}
  {2011})}\BibitemShut {NoStop}%
\bibitem [{\citenamefont {Siddons}\ \emph {et~al.}(2008)\citenamefont
  {Siddons}, \citenamefont {Adams}, \citenamefont {Ge},\ and\ \citenamefont
  {Hughes}}]{Siddons2008}%
  \BibitemOpen
  \bibfield  {author} {\bibinfo {author} {\bibfnamefont {P.}~\bibnamefont
  {Siddons}}, \bibinfo {author} {\bibfnamefont {C.~S.}\ \bibnamefont {Adams}},
  \bibinfo {author} {\bibfnamefont {C.}~\bibnamefont {Ge}},\ and\ \bibinfo
  {author} {\bibfnamefont {I.~G.}\ \bibnamefont {Hughes}},\ }\bibfield  {title}
  {\bibinfo {title} {{Absolute absorption on rubidium D lines: comparison
  between theory and experiment}},\ }\href
  {https://doi.org/10.1088/0953-4075/41/15/155004} {\bibfield  {journal}
  {\bibinfo  {journal} {J. Phys. B: At. Mol. Opt. Phys.}\ }\textbf {\bibinfo
  {volume} {41}},\ \bibinfo {pages} {155004} (\bibinfo {year}
  {2008})}\BibitemShut {NoStop}%
\bibitem [{\citenamefont {Weller}\ \emph {et~al.}(2011)\citenamefont {Weller},
  \citenamefont {Bettles}, \citenamefont {Siddons}, \citenamefont {Adams},\
  and\ \citenamefont {Hughes}}]{Weller2011}%
  \BibitemOpen
  \bibfield  {author} {\bibinfo {author} {\bibfnamefont {L.}~\bibnamefont
  {Weller}}, \bibinfo {author} {\bibfnamefont {R.~J.}\ \bibnamefont {Bettles}},
  \bibinfo {author} {\bibfnamefont {P.}~\bibnamefont {Siddons}}, \bibinfo
  {author} {\bibfnamefont {C.~S.}\ \bibnamefont {Adams}},\ and\ \bibinfo
  {author} {\bibfnamefont {I.~G.}\ \bibnamefont {Hughes}},\ }\bibfield  {title}
  {\bibinfo {title} {{Absolute absorption on the rubidium D$_1$ line including
  resonant dipole–dipole interactions}},\ }\href
  {https://doi.org/10.1088/0953-4075/44/19/195006} {\bibfield  {journal}
  {\bibinfo  {journal} {J. Phys. B: At. Mol. Opt. Phys.}\ }\textbf {\bibinfo
  {volume} {44}},\ \bibinfo {pages} {195006} (\bibinfo {year}
  {2011})}\BibitemShut {NoStop}%
\bibitem [{\citenamefont {Zentile}\ \emph
  {et~al.}(2015{\natexlab{a}})\citenamefont {Zentile}, \citenamefont
  {Keaveney}, \citenamefont {Weller}, \citenamefont {Whiting}, \citenamefont
  {Adams},\ and\ \citenamefont {Hughes}}]{Zentile2014a}%
  \BibitemOpen
  \bibfield  {author} {\bibinfo {author} {\bibfnamefont {M.~A.}\ \bibnamefont
  {Zentile}}, \bibinfo {author} {\bibfnamefont {J.}~\bibnamefont {Keaveney}},
  \bibinfo {author} {\bibfnamefont {L.}~\bibnamefont {Weller}}, \bibinfo
  {author} {\bibfnamefont {D.~J.}\ \bibnamefont {Whiting}}, \bibinfo {author}
  {\bibfnamefont {C.~S.}\ \bibnamefont {Adams}},\ and\ \bibinfo {author}
  {\bibfnamefont {I.~G.}\ \bibnamefont {Hughes}},\ }\bibfield  {title}
  {\bibinfo {title} {{ElecSus: A program to calculate the electric
  susceptibility of an atomic ensemble}},\ }\href
  {https://doi.org/10.1016/j.cpc.2014.11.023} {\bibfield  {journal} {\bibinfo
  {journal} {Comput. Phys. Commun.}\ }\textbf {\bibinfo {volume} {189}},\
  \bibinfo {pages} {162} (\bibinfo {year} {2015}{\natexlab{a}})}\BibitemShut
  {NoStop}%
\bibitem [{\citenamefont {Keaveney}\ \emph
  {et~al.}(2018{\natexlab{a}})\citenamefont {Keaveney}, \citenamefont {Adams},\
  and\ \citenamefont {Hughes}}]{Keaveney2018}%
  \BibitemOpen
  \bibfield  {author} {\bibinfo {author} {\bibfnamefont {J.}~\bibnamefont
  {Keaveney}}, \bibinfo {author} {\bibfnamefont {C.~S.}\ \bibnamefont
  {Adams}},\ and\ \bibinfo {author} {\bibfnamefont {I.~G.}\ \bibnamefont
  {Hughes}},\ }\bibfield  {title} {\bibinfo {title} {{ElecSus: Extension to
  arbitrary geometry magneto-optics}},\ }\href
  {https://doi.org/10.1016/j.cpc.2017.12.001} {\bibfield  {journal} {\bibinfo
  {journal} {Comput. Phys. Commun.}\ }\textbf {\bibinfo {volume} {224}},\
  \bibinfo {pages} {311} (\bibinfo {year} {2018}{\natexlab{a}})}\BibitemShut
  {NoStop}%
\bibitem [{\citenamefont {Kiefer}\ \emph {et~al.}(2014)\citenamefont {Kiefer},
  \citenamefont {L{\"{o}}w}, \citenamefont {Wrachtrup},\ and\ \citenamefont
  {Gerhardt}}]{Kiefer2014}%
  \BibitemOpen
  \bibfield  {author} {\bibinfo {author} {\bibfnamefont {W.}~\bibnamefont
  {Kiefer}}, \bibinfo {author} {\bibfnamefont {R.}~\bibnamefont {L{\"{o}}w}},
  \bibinfo {author} {\bibfnamefont {J.}~\bibnamefont {Wrachtrup}},\ and\
  \bibinfo {author} {\bibfnamefont {I.}~\bibnamefont {Gerhardt}},\ }\bibfield
  {title} {\bibinfo {title} {{Na-Faraday rotation filtering: The optimal
  point}},\ }\href {https://doi.org/10.1038/srep06552} {\bibfield  {journal}
  {\bibinfo  {journal} {Sci. Rep.}\ }\textbf {\bibinfo {volume} {4}},\ \bibinfo
  {pages} {6552} (\bibinfo {year} {2014})}\BibitemShut {NoStop}%
\bibitem [{\citenamefont {Zentile}\ \emph
  {et~al.}(2015{\natexlab{b}})\citenamefont {Zentile}, \citenamefont
  {Keaveney}, \citenamefont {Mathew}, \citenamefont {Whiting}, \citenamefont
  {Adams},\ and\ \citenamefont {Hughes}}]{Zentile2015}%
  \BibitemOpen
  \bibfield  {author} {\bibinfo {author} {\bibfnamefont {M.~A.}\ \bibnamefont
  {Zentile}}, \bibinfo {author} {\bibfnamefont {J.}~\bibnamefont {Keaveney}},
  \bibinfo {author} {\bibfnamefont {R.~S.}\ \bibnamefont {Mathew}}, \bibinfo
  {author} {\bibfnamefont {D.~J.}\ \bibnamefont {Whiting}}, \bibinfo {author}
  {\bibfnamefont {C.~S.}\ \bibnamefont {Adams}},\ and\ \bibinfo {author}
  {\bibfnamefont {I.~G.}\ \bibnamefont {Hughes}},\ }\bibfield  {title}
  {\bibinfo {title} {{Optimization of atomic Faraday filters in the presence of
  homogeneous line broadening}},\ }\href
  {https://doi.org/10.1088/0953-4075/48/18/185001} {\bibfield  {journal}
  {\bibinfo  {journal} {J. Phys. B: At. Mol. Opt. Phys.}\ }\textbf {\bibinfo
  {volume} {48}},\ \bibinfo {pages} {185001} (\bibinfo {year}
  {2015}{\natexlab{b}})}\BibitemShut {NoStop}%
\bibitem [{\citenamefont {Keaveney}\ \emph
  {et~al.}(2018{\natexlab{b}})\citenamefont {Keaveney}, \citenamefont
  {Wrathmall}, \citenamefont {Adams},\ and\ \citenamefont
  {Hughes}}]{Keaveney2018d}%
  \BibitemOpen
  \bibfield  {author} {\bibinfo {author} {\bibfnamefont {J.}~\bibnamefont
  {Keaveney}}, \bibinfo {author} {\bibfnamefont {S.~A.}\ \bibnamefont
  {Wrathmall}}, \bibinfo {author} {\bibfnamefont {C.~S.}\ \bibnamefont
  {Adams}},\ and\ \bibinfo {author} {\bibfnamefont {I.~G.}\ \bibnamefont
  {Hughes}},\ }\bibfield  {title} {\bibinfo {title} {{Optimized ultra-narrow
  atomic bandpass filters via magneto-optic rotation in an unconstrained
  geometry}},\ }\href {https://doi.org/10.1364/OL.43.004272} {\bibfield
  {journal} {\bibinfo  {journal} {Opt. Lett.}\ }\textbf {\bibinfo {volume}
  {43}},\ \bibinfo {pages} {4272} (\bibinfo {year}
  {2018}{\natexlab{b}})}\BibitemShut {NoStop}%
\bibitem [{\citenamefont {Reed}\ \emph {et~al.}(2018)\citenamefont {Reed},
  \citenamefont {{\v{S}}ibali{\'{c}}}, \citenamefont {Whiting}, \citenamefont
  {Kondo}, \citenamefont {Adams},\ and\ \citenamefont {Weatherill}}]{Reed2018}%
  \BibitemOpen
  \bibfield  {author} {\bibinfo {author} {\bibfnamefont {D.~J.}\ \bibnamefont
  {Reed}}, \bibinfo {author} {\bibfnamefont {N.}~\bibnamefont
  {{\v{S}}ibali{\'{c}}}}, \bibinfo {author} {\bibfnamefont {D.~J.}\
  \bibnamefont {Whiting}}, \bibinfo {author} {\bibfnamefont {J.~M.}\
  \bibnamefont {Kondo}}, \bibinfo {author} {\bibfnamefont {C.~S.}\ \bibnamefont
  {Adams}},\ and\ \bibinfo {author} {\bibfnamefont {K.~J.}\ \bibnamefont
  {Weatherill}},\ }\bibfield  {title} {\bibinfo {title} {{Low-drift Zeeman
  shifted atomic frequency reference}},\ }\href
  {https://doi.org/10.1364/OSAC.1.000004} {\bibfield  {journal} {\bibinfo
  {journal} {OSA Continuum}\ }\textbf {\bibinfo {volume} {1}},\ \bibinfo
  {pages} {4} (\bibinfo {year} {2018})}\BibitemShut {NoStop}%
\bibitem [{\citenamefont {Sargsyan}\ \emph {et~al.}(2012)\citenamefont
  {Sargsyan}, \citenamefont {Hakhumyan}, \citenamefont {Leroy}, \citenamefont
  {Pashayan-Leroy}, \citenamefont {Papoyan},\ and\ \citenamefont
  {Sarkisyan}}]{Sargsyan2012a}%
  \BibitemOpen
  \bibfield  {author} {\bibinfo {author} {\bibfnamefont {A.}~\bibnamefont
  {Sargsyan}}, \bibinfo {author} {\bibfnamefont {G.}~\bibnamefont {Hakhumyan}},
  \bibinfo {author} {\bibfnamefont {C.}~\bibnamefont {Leroy}}, \bibinfo
  {author} {\bibfnamefont {Y.}~\bibnamefont {Pashayan-Leroy}}, \bibinfo
  {author} {\bibfnamefont {A.}~\bibnamefont {Papoyan}},\ and\ \bibinfo {author}
  {\bibfnamefont {D.}~\bibnamefont {Sarkisyan}},\ }\bibfield  {title} {\bibinfo
  {title} {{Hyperfine Paschen-Back regime realized in Rb nanocell}},\ }\href
  {https://doi.org/10.1364/OL.37.001379} {\bibfield  {journal} {\bibinfo
  {journal} {Opt. Lett.}\ }\textbf {\bibinfo {volume} {37}},\ \bibinfo {pages}
  {1379} (\bibinfo {year} {2012})}\BibitemShut {NoStop}%
\bibitem [{\citenamefont {Weller}\ \emph
  {et~al.}(2012{\natexlab{c}})\citenamefont {Weller}, \citenamefont
  {Kleinbach}, \citenamefont {Zentile}, \citenamefont {Knappe}, \citenamefont
  {Adams},\ and\ \citenamefont {Hughes}}]{Weller2012c}%
  \BibitemOpen
  \bibfield  {author} {\bibinfo {author} {\bibfnamefont {L.}~\bibnamefont
  {Weller}}, \bibinfo {author} {\bibfnamefont {K.~S.}\ \bibnamefont
  {Kleinbach}}, \bibinfo {author} {\bibfnamefont {M.~A.}\ \bibnamefont
  {Zentile}}, \bibinfo {author} {\bibfnamefont {S.}~\bibnamefont {Knappe}},
  \bibinfo {author} {\bibfnamefont {C.~S.}\ \bibnamefont {Adams}},\ and\
  \bibinfo {author} {\bibfnamefont {I.~G.}\ \bibnamefont {Hughes}},\ }\bibfield
   {title} {\bibinfo {title} {{Absolute absorption and dispersion of a rubidium
  vapour in the hyperfine Paschen–Back regime}},\ }\href
  {https://doi.org/10.1088/0953-4075/45/21/215005} {\bibfield  {journal}
  {\bibinfo  {journal} {J. Phys. B: At. Mol. Opt. Phys.}\ }\textbf {\bibinfo
  {volume} {45}},\ \bibinfo {pages} {215005} (\bibinfo {year}
  {2012}{\natexlab{c}})}\BibitemShut {NoStop}%
\bibitem [{\citenamefont {Zentile}\ \emph {et~al.}(2014)\citenamefont
  {Zentile}, \citenamefont {Andrews}, \citenamefont {Weller}, \citenamefont
  {Knappe}, \citenamefont {Adams},\ and\ \citenamefont {Hughes}}]{Zentile2014}%
  \BibitemOpen
  \bibfield  {author} {\bibinfo {author} {\bibfnamefont {M.~A.}\ \bibnamefont
  {Zentile}}, \bibinfo {author} {\bibfnamefont {R.}~\bibnamefont {Andrews}},
  \bibinfo {author} {\bibfnamefont {L.}~\bibnamefont {Weller}}, \bibinfo
  {author} {\bibfnamefont {S.}~\bibnamefont {Knappe}}, \bibinfo {author}
  {\bibfnamefont {C.~S.}\ \bibnamefont {Adams}},\ and\ \bibinfo {author}
  {\bibfnamefont {I.~G.}\ \bibnamefont {Hughes}},\ }\bibfield  {title}
  {\bibinfo {title} {{The hyperfine Paschen–Back Faraday effect}},\ }\href
  {https://doi.org/10.1088/0953-4075/47/7/075005} {\bibfield  {journal}
  {\bibinfo  {journal} {J. Phys. B: At. Mol. Opt. Phys.}\ }\textbf {\bibinfo
  {volume} {47}},\ \bibinfo {pages} {075005} (\bibinfo {year}
  {2014})}\BibitemShut {NoStop}%
\bibitem [{\citenamefont {Sargsyan}\ \emph {et~al.}(2015)\citenamefont
  {Sargsyan}, \citenamefont {Tonoyan}, \citenamefont {Hakhumyan}, \citenamefont
  {Leroy}, \citenamefont {Pashayan-Leroy},\ and\ \citenamefont
  {Sarkisyan}}]{Sargsyan2015a}%
  \BibitemOpen
  \bibfield  {author} {\bibinfo {author} {\bibfnamefont {A.}~\bibnamefont
  {Sargsyan}}, \bibinfo {author} {\bibfnamefont {A.}~\bibnamefont {Tonoyan}},
  \bibinfo {author} {\bibfnamefont {G.}~\bibnamefont {Hakhumyan}}, \bibinfo
  {author} {\bibfnamefont {C.}~\bibnamefont {Leroy}}, \bibinfo {author}
  {\bibfnamefont {Y.}~\bibnamefont {Pashayan-Leroy}},\ and\ \bibinfo {author}
  {\bibfnamefont {D.}~\bibnamefont {Sarkisyan}},\ }\bibfield  {title} {\bibinfo
  {title} {{Complete hyperfine Paschen-Back regime at relatively small magnetic
  fields realized in potassium nano-cell}},\ }\href
  {https://doi.org/10.1209/0295-5075/110/23001} {\bibfield  {journal} {\bibinfo
   {journal} {EPL}\ }\textbf {\bibinfo {volume} {110}},\ \bibinfo {pages}
  {23001} (\bibinfo {year} {2015})}\BibitemShut {NoStop}%
\bibitem [{\citenamefont {Epple}\ \emph {et~al.}(2014)\citenamefont {Epple},
  \citenamefont {Kleinbach}, \citenamefont {Euser}, \citenamefont {Joly},
  \citenamefont {Pfau}, \citenamefont {Russell},\ and\ \citenamefont
  {L{\"{o}}w}}]{Epple2014}%
  \BibitemOpen
  \bibfield  {author} {\bibinfo {author} {\bibfnamefont {G.}~\bibnamefont
  {Epple}}, \bibinfo {author} {\bibfnamefont {K.~S.}\ \bibnamefont
  {Kleinbach}}, \bibinfo {author} {\bibfnamefont {T.~G.}\ \bibnamefont
  {Euser}}, \bibinfo {author} {\bibfnamefont {N.~Y.}\ \bibnamefont {Joly}},
  \bibinfo {author} {\bibfnamefont {T.}~\bibnamefont {Pfau}}, \bibinfo {author}
  {\bibfnamefont {P.~{\relax St}.~J.}\ \bibnamefont {Russell}},\ and\ \bibinfo
  {author} {\bibfnamefont {R.}~\bibnamefont {L{\"{o}}w}},\ }\bibfield  {title}
  {\bibinfo {title} {{Rydberg atoms in hollow-core photonic crystal fibres}},\
  }\href {https://doi.org/10.1038/ncomms5132} {\bibfield  {journal} {\bibinfo
  {journal} {Nat. Commun.}\ }\textbf {\bibinfo {volume} {5}},\ \bibinfo {pages}
  {4132} (\bibinfo {year} {2014})}\BibitemShut {NoStop}%
\bibitem [{\citenamefont {Gutekunst}(2016)}]{Gutekunst2016}%
  \BibitemOpen
  \bibfield  {author} {\bibinfo {author} {\bibfnamefont {J.}~\bibnamefont
  {Gutekunst}},\ }\emph {\bibinfo {title} {{EIT Spectroscopy in Hollow Core
  Fibers}}},\ \href@noop {} {\bibinfo {type} {Master thesis}},\ \bibinfo
  {school} {University of Stuttgart} (\bibinfo {year} {2016})\BibitemShut
  {NoStop}%
\bibitem [{\citenamefont {Gutekunst}\ \emph {et~al.}(2017)\citenamefont
  {Gutekunst}, \citenamefont {Weller}, \citenamefont {K{\"{u}}bler},
  \citenamefont {Negel}, \citenamefont {Ahmed}, \citenamefont {Graf},\ and\
  \citenamefont {L{\"{o}}w}}]{Gutekunst2017}%
  \BibitemOpen
  \bibfield  {author} {\bibinfo {author} {\bibfnamefont {J.}~\bibnamefont
  {Gutekunst}}, \bibinfo {author} {\bibfnamefont {D.}~\bibnamefont {Weller}},
  \bibinfo {author} {\bibfnamefont {H.}~\bibnamefont {K{\"{u}}bler}}, \bibinfo
  {author} {\bibfnamefont {J.-P.}\ \bibnamefont {Negel}}, \bibinfo {author}
  {\bibfnamefont {M.~A.}\ \bibnamefont {Ahmed}}, \bibinfo {author}
  {\bibfnamefont {T.}~\bibnamefont {Graf}},\ and\ \bibinfo {author}
  {\bibfnamefont {R.}~\bibnamefont {L{\"{o}}w}},\ }\bibfield  {title} {\bibinfo
  {title} {{Fiber-integrated spectroscopy device for hot alkali vapor}},\
  }\href {https://doi.org/10.1364/AO.56.005898} {\bibfield  {journal} {\bibinfo
   {journal} {Appl. Opt.}\ }\textbf {\bibinfo {volume} {56}},\ \bibinfo {pages}
  {5898} (\bibinfo {year} {2017})}\BibitemShut {NoStop}%
\bibitem [{\citenamefont {Kaczmarek}\ \emph {et~al.}(2015)\citenamefont
  {Kaczmarek}, \citenamefont {Saunders}, \citenamefont {Sprague}, \citenamefont
  {Kolthammer}, \citenamefont {Feizpour}, \citenamefont {Ledingham},
  \citenamefont {Brecht}, \citenamefont {Poem}, \citenamefont {Walmsley},\ and\
  \citenamefont {Nunn}}]{Kaczmarek2015a}%
  \BibitemOpen
  \bibfield  {author} {\bibinfo {author} {\bibfnamefont {K.~T.}\ \bibnamefont
  {Kaczmarek}}, \bibinfo {author} {\bibfnamefont {D.~J.}\ \bibnamefont
  {Saunders}}, \bibinfo {author} {\bibfnamefont {M.~R.}\ \bibnamefont
  {Sprague}}, \bibinfo {author} {\bibfnamefont {W.~S.}\ \bibnamefont
  {Kolthammer}}, \bibinfo {author} {\bibfnamefont {A.}~\bibnamefont
  {Feizpour}}, \bibinfo {author} {\bibfnamefont {P.~M.}\ \bibnamefont
  {Ledingham}}, \bibinfo {author} {\bibfnamefont {B.}~\bibnamefont {Brecht}},
  \bibinfo {author} {\bibfnamefont {E.}~\bibnamefont {Poem}}, \bibinfo {author}
  {\bibfnamefont {I.~A.}\ \bibnamefont {Walmsley}},\ and\ \bibinfo {author}
  {\bibfnamefont {J.}~\bibnamefont {Nunn}},\ }\bibfield  {title} {\bibinfo
  {title} {{Ultrahigh and persistent optical depths of cesium in
  Kagom{\'{e}}-type hollow-core photonic crystal fibers}},\ }\href
  {https://doi.org/10.1364/OL.40.005582} {\bibfield  {journal} {\bibinfo
  {journal} {Opt. Lett.}\ }\textbf {\bibinfo {volume} {40}},\ \bibinfo {pages}
  {5582} (\bibinfo {year} {2015})}\BibitemShut {NoStop}%
\bibitem [{\citenamefont {Donvalkar}\ \emph {et~al.}(2015)\citenamefont
  {Donvalkar}, \citenamefont {Ramelow}, \citenamefont {Clemmen},\ and\
  \citenamefont {Gaeta}}]{Donvalkar2015a}%
  \BibitemOpen
  \bibfield  {author} {\bibinfo {author} {\bibfnamefont {P.~S.}\ \bibnamefont
  {Donvalkar}}, \bibinfo {author} {\bibfnamefont {S.}~\bibnamefont {Ramelow}},
  \bibinfo {author} {\bibfnamefont {S.}~\bibnamefont {Clemmen}},\ and\ \bibinfo
  {author} {\bibfnamefont {A.~L.}\ \bibnamefont {Gaeta}},\ }\bibfield  {title}
  {\bibinfo {title} {{Continuous generation of rubidium vapor in hollow-core
  photonic bandgap fibers}},\ }\href {https://doi.org/10.1364/OL.40.005379}
  {\bibfield  {journal} {\bibinfo  {journal} {Opt. Lett.}\ }\textbf {\bibinfo
  {volume} {40}},\ \bibinfo {pages} {5379} (\bibinfo {year}
  {2015})}\BibitemShut {NoStop}%
\bibitem [{\citenamefont {Pooser}\ \emph {et~al.}(2009)\citenamefont {Pooser},
  \citenamefont {Marino}, \citenamefont {Boyer}, \citenamefont {Jones},\ and\
  \citenamefont {Lett}}]{Pooser2009}%
  \BibitemOpen
  \bibfield  {author} {\bibinfo {author} {\bibfnamefont {R.~C.}\ \bibnamefont
  {Pooser}}, \bibinfo {author} {\bibfnamefont {A.~M.}\ \bibnamefont {Marino}},
  \bibinfo {author} {\bibfnamefont {V.}~\bibnamefont {Boyer}}, \bibinfo
  {author} {\bibfnamefont {K.~M.}\ \bibnamefont {Jones}},\ and\ \bibinfo
  {author} {\bibfnamefont {P.~D.}\ \bibnamefont {Lett}},\ }\bibfield  {title}
  {\bibinfo {title} {{Quantum correlated light beams from non-degenerate
  four-wave mixing in an atomic vapor: the D1 and D2 lines of $^{85}$Rb and
  $^{87}$Rb}},\ }\href {https://doi.org/10.1364/OE.17.016722} {\bibfield
  {journal} {\bibinfo  {journal} {Opt. Express}\ }\textbf {\bibinfo {volume}
  {17}},\ \bibinfo {pages} {16722} (\bibinfo {year} {2009})}\BibitemShut
  {NoStop}%
\bibitem [{\citenamefont {Cohen-Tannoudji}(1996)}]{Cohen-Tannoudji1996}%
  \BibitemOpen
  \bibfield  {author} {\bibinfo {author} {\bibfnamefont {C.~N.}\ \bibnamefont
  {Cohen-Tannoudji}},\ }\bibfield  {title} {\bibinfo {title} {{The
  Autler-Townes Effect Revisited}},\ }in\ \href
  {https://doi.org/10.1007/978-1-4612-2378-8_11} {\emph {\bibinfo {booktitle}
  {Amazing Light}}}\ (\bibinfo  {publisher} {Springer},\ \bibinfo {address}
  {New York, NY},\ \bibinfo {year} {1996})\ pp.\ \bibinfo {pages}
  {109--123}\BibitemShut {NoStop}%
\bibitem [{\citenamefont {Bradley}(2013)}]{Version2019}%
  \BibitemOpen
  \bibfield  {author} {\bibinfo {author} {\bibfnamefont {T.~D.}\ \bibnamefont
  {Bradley}},\ }\emph {\bibinfo {title} {{Atomic vapours filled hollow core
  photonic crystal fibre for magneto-optical spectroscopy}}},\ \href@noop {}
  {Ph.D. thesis},\ \bibinfo  {school} {University of Bath} (\bibinfo {year}
  {2013})\BibitemShut {NoStop}%
\bibitem [{\citenamefont {Brandt}\ \emph {et~al.}(1997)\citenamefont {Brandt},
  \citenamefont {Nagel}, \citenamefont {Wynands},\ and\ \citenamefont
  {Meschede}}]{Brandt1997}%
  \BibitemOpen
  \bibfield  {author} {\bibinfo {author} {\bibfnamefont {S.}~\bibnamefont
  {Brandt}}, \bibinfo {author} {\bibfnamefont {A.}~\bibnamefont {Nagel}},
  \bibinfo {author} {\bibfnamefont {R.}~\bibnamefont {Wynands}},\ and\ \bibinfo
  {author} {\bibfnamefont {D.}~\bibnamefont {Meschede}},\ }\bibfield  {title}
  {\bibinfo {title} {{Buffer-gas-induced linewidth reduction of coherent dark
  resonances to below 50 Hz}},\ }\href
  {https://doi.org/10.1103/PhysRevA.56.R1063} {\bibfield  {journal} {\bibinfo
  {journal} {Phys. Rev. A}\ }\textbf {\bibinfo {volume} {56}},\ \bibinfo
  {pages} {R1063} (\bibinfo {year} {1997})}\BibitemShut {NoStop}%
\end{thebibliography}%

\end{document}